\documentclass[aps,prb,amsfonts,amssymb,twocolumn,groupedaddress]{revtex4}

\usepackage{graphicx}

\newcommand{\greeksym}[1]{{\usefont{U}{psy}{m}{n}#1}}
\newcommand{\usigma}{\mbox{\greeksym{s}}}

\begin{document}

\title{Superconducting tetrahedral quantum bits}

\author{M.V.\ Feigel'man$^{\, 1}$, L.B.\ Ioffe$^{\, 2,1}$, V.B.\
Geshkenbein$^{\, 3,1}$, P.\ Dayal$^{\, 3}$, and G.\
Blatter$^{\, 3}$}

\affiliation{$^1$Landau Institute for Theoretical Physics, 117940 Moscow,
Russia}
\affiliation{$^2$Department of Physics and Astronomy, Rutgers University,
Piscataway, NJ 08854, USA}
\affiliation{$^3$Theoretische Physik, ETH-H\"onggerberg, CH-8093 Z\"urich,
Switzerland}

\date{\today}

\begin{abstract}
 We propose a new design for a quantum bit with four superconducting
 islands in the topology of a symmetric tetrahedron, uniformly
 frustrated with one-half flux-quantum per loop and one-half
 Cooper-pair per island. This structure emulates a
 noise-resistant spin-1/2 system in a vanishing magnetic field.
 The tetrahedral quantum bit combines a number of advances such as
 a doubly degeneracy ground state minimizing decoherence via phonon
 radiation, a weak quadratic sensitivity to electric and
 magnetic noise, relieved constraints on the junction fabrication,
 a large freedom in manipulation, and attractive measurement schemes.
 The simultaneous appearance of a degenerate ground state and a
 weak noise sensitivity are consequences of the tetrahedral
 symmetry, while enhanced quantum fluctuations derive from
 the special magnetic frustration. We determine the spectral
 properties of the tetrahedral structure within a semiclassical
 analysis and confirm the results numerically. We show how
 proper tuning of the charge-frustration selects a doubly degenerate
 ground state and discuss the qubit's manipulation through
 capacitive and inductive coupling to external bias sources.
 The complete readout of the spin-components $\sigma_i$,
 $i=x,y,z$, is achieved through coupling of the internal
 qubit currents to external junctions driven close to
 criticality during the measurement.
\end{abstract}

\maketitle

\section{Introduction}
Superconducting solid-state qubits (short for quantum bits) are
promising candidates for the future construction of quantum
information processors. They appear in a variety of designs:
in the charge-version \cite{nakamura_99,shnirman_97,averin_98}
the quantum information is stored in the number of excess
Cooper pairs residing on a small superconducting island
--- this design requires fabrication of ultra-small structures
and is susceptible to charge noise. In the flux- \cite{friedman_00}
and phase- \cite{vanderwal_00,chiorescu_03} versions the
information is encoded in the current state of the device ---
this is a macroscopic variable susceptible to flux noise. The new
design by Vion {\it et al.} \cite{vion_02} is `in between', with
the energy scales for the charge- ($E_C = e^2 /2C$) and the phase-
($E_J = \Phi_0 I_c/2\pi c$) degrees of freedom roughly balancing
one another (here, $C$ and $I_c$ are the capacitance and the
maximal current of the device, $\Phi_0 = hc/2e$ is the flux
quantum); its ground state is non-degenerate and the limitations
are close to those of the charge device.

The novel tetrahedral qubit design we propose below operates in
the phase-dominated regime and exhibits two remarkable physical
properties: first, its non-Abelian symmetry group (the tetrahedral
group $T_d$) leads to the natural appearance of degenerate states
and appropriate tuning of parameters provides us with a doubly
degenerate groundstate. Our tetrahedral qubit then emulates a
spin-1/2 system in a vanishing magnetic field, the ideal starting
point for the construction of a qubit. Manipulation of the
tetrahedral qubit through external bias signals translates into
application of magnetic fields on the spin; the application of the
bias to different elements of the tetrahedral qubit corresponds to
rotated operations in spin space. Furthermore, geometric quantum
computation via Berry phases \cite{jones_00,duan_01} might be
implemented through adiabatic change of external variables. Going
one step further, one may hope to make use of this type of systems
in the future physical realization of non-Abelian anyons, thereby
aiming at a new generation of topological devices
\cite{ioffe1_02,ioffe2_02,doucot1_02,doucot2_02} which keep their
protection even during operation \cite{kitaev_97}.
\begin{figure}[ht]
\includegraphics[scale=0.23]{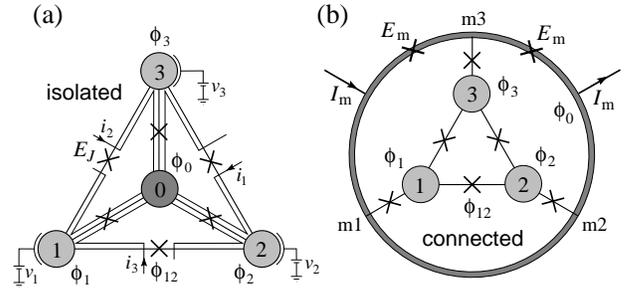}
\caption[]{(a) Tetrahedral superconducting qubit involving four
  islands and six junctions (with Josephson coupling $E_J$
  and charging energy $E_C$); all islands and junctions are assumed
  to be equal and arranged in a symmetric way. The islands are
  attributed phases $\phi_i$, $i=0,\dots,3$. The qubit is
  manipulated via bias voltages $v_i$ and bias currents $i_i$.
  In order to measure the qubit's state it is convenient to
  invert the tetrahedron as shown in (b) --- we refer to
  this version as the `connected' tetrahedron with the
  inner dark-grey island in (a) transformed into the outer
  ring in (b). The measurement involves additional measurement
  junctions with couplings $E_{\rm m} \gg E_J$ on the outer ring
  which are driven by external currents $I_{\rm m}$ (schematic,
  see Fig.\ 6 for details); the large coupling $E_{\rm m}$
  effectively binds the ring segments into one island.}
  \label{fig:tetrahedron_1}
\end{figure}

The second property we wish to exploit is geometric frustration:
In our tetrahedral qubit discussed below it appears in an extreme
way by rendering the classical minimal states continuously
degenerate along a line in parameter space. Semi-classical states
then appear only through a fluctuation-induced potential,
reminiscent of the Casimir effect \cite{casimir_48} and the
concept of inducing `order from disorder'
\cite{villain_80,tsvelik_95}. The quantum-tunneling between these
semi-classical states defines the operational energy scale of the
qubit, which turns out to be unusually large due to the weakness
of the fluctuation-induced potential. Hence the geometric
frustration present in our tetrahedral qubit provides a natural
boost for the quantum fluctuations without the stringent
requirements on the smallness of the junction capacitances, thus
avoiding the disadvantages of both the charge- and the
phase-device: The larger junctions reduce the demands on the
fabrication process and the susceptibility to charge noise and
mesoscopic effects, while the large operational energy scale due
to the soft fluctuation-induced potential reduces the effects of
flux noise. Both types of electromagnetic noise, charge- and flux
noise, appear only in second order (cf.\ also Ref.\
\onlinecite{vion_02}). 

In the following (Sec.\ 2), we first introduce the structure of our
tetrahedral qubit and then find the low-energy part of its spectrum.
We proceed in three steps, beginning with the (highly degenerate) 
classical solution; subsequently, we demonstrate how the fluctuation
induced potential reduces this degeneracy to three semi-classical 
states and finally, we analyze the tunneling between them in order to
arrive at the final answer for the phase dominated regime. We confirm 
this solution with the help of numerical calculations and extend 
it to the charge dominated regime. The inclusion of external fields
breaking the (tetrahedral) symmetry of the device prepares the 
discussion of the qubit's manipulation schemes (section 3). In section
4, we discuss various measurement schemes and end with the conclusions
in Sec.\ 5.  A first account on part of this work has been given in 
Ref.\ \onlinecite{tetra_short}.

\section{Tetrahedron}
\subsection{Device structure}
Consider the planar structure made from four superconducting
islands interconnected via six (conventional) Josephson junctions
with equal couplings $E_J$; connecting each island with all the
others produces the topology of a tetrahedron, see Fig.\ 1(a). The
islands are numbered through `0' to `3', with the island `0'
residing in the center, and are assigned phases $\phi_i$,
$i=0,\dots 3$. The three triangular loops are small (i.e., $E_J
\ll (\Phi_0/2\pi)^2/L_\triangle$ with $L_\triangle$ the loop
inductance), allowing us to neglect fluxes induced by currents
flowing in the structure. In the absence of external magnetic
fields, the classical energy of this arrangement is given by the
sum $V_0 = \sum_{i<j}E_J\, [1-\cos(\phi_{ij})]$ with the
difference variables $\phi_{ij}= \phi_j-\phi_i$. A slightly
modified version of this device with the inner island converted
into an enclosing ring is shown in Fig.\ 1(b) --- choosing
appropriate parameters, this variant exhibits the same physical
properties as the original isolated tetrahedral qubit. In
addition, this second design ideally lends itself for measurement
of the qubit state. Below, we treat the ring as one connected
island; the strong junctions with $E_{\rm m} \gg E_J$ used in the
measurement process will be discussed later.

Next, we bias the structure through an external magnetic field,
frustrating each sub-loop with a flux $\Phi_0/2$ (these are three
triangular sub-loops in the isolated version of Fig.\ 1(a) and 4
sub-loops in the connected version of Fig.\ 1(b)). We include the
effect of this flux in a symmetric gauge by adding the phase $\pi$
to each of the difference variables $\phi_{ij}$; the energy (up to
a trivial constant; we measure phases with respect to the phase
$\phi_0$ of the central/ring island)
\begin{eqnarray}
   V_\pi &=& E_J [\cos\phi_1+\cos\phi_2+\cos\phi_3
   \label{EJ}\\ &&\quad
   +\cos(\phi_3-\phi_2)+\cos(\phi_1-\phi_3)+\cos(\phi_2-\phi_1)]
   \nonumber
\end{eqnarray}
then is minimized (with $V_\pi = -2E_J$) along the lines
\begin{equation}
   \phi_3=\pm\pi;\,\,\, \phi_2-\phi_1 = \pm \pi;\,\,\,
   {\psi}_3 = \phi_1+\phi_2 \in [-\pi,\pi]
   \label{lines}
\end{equation}
and their analogs obtained by cyclic permutation $3 \rightarrow 1
\rightarrow 2 \rightarrow 3$. These minimal-energy lines run in
the planes of the cube $[-\pi,\pi]^3$ defined in phase space
$\{\phi_1,\phi_2,\phi_3\}$, see Fig.\ 2(a). The huge (linear)
classical degeneracy can be easily understood via a reformulation
of the potential (\ref{EJ}) in terms of the complex variables $z_k
= \exp(i\phi_k)$, $2V_\pi = E_J [|\sum_{k=0}^3 z_k|^2-4]$; this
expression is minimal for $\sum_k z_k = 0$. The two conditions
defined by this (complex) equation imply that one out of the three
variables $\phi_k$ can be freely chosen, thus defining lines of
minimal potential energy. Of particular relevance are the minimal
energy configurations $O_i$, $i=1,2,3$ on the cube edges where two
minimal-energy lines join. These configurations involve two
opposite junctions with a phase difference $\phi_{ij}=0$, while
the remaining junctions involve maximal phase differences $\pi$;
following three consecutive segments on the cube, these minimal
states rotate through $2\pi$, see Fig.\ 2(a) (here, we have
included the bias phases $\pi$ on each link; in the original
variables these states involve 2 strained junctions with phase
difference $\pi$ and 4 unstrained ones with phase difference 0;
note the absence of currents in these configurations).

\begin{figure}[ht]
\includegraphics[scale=0.3]{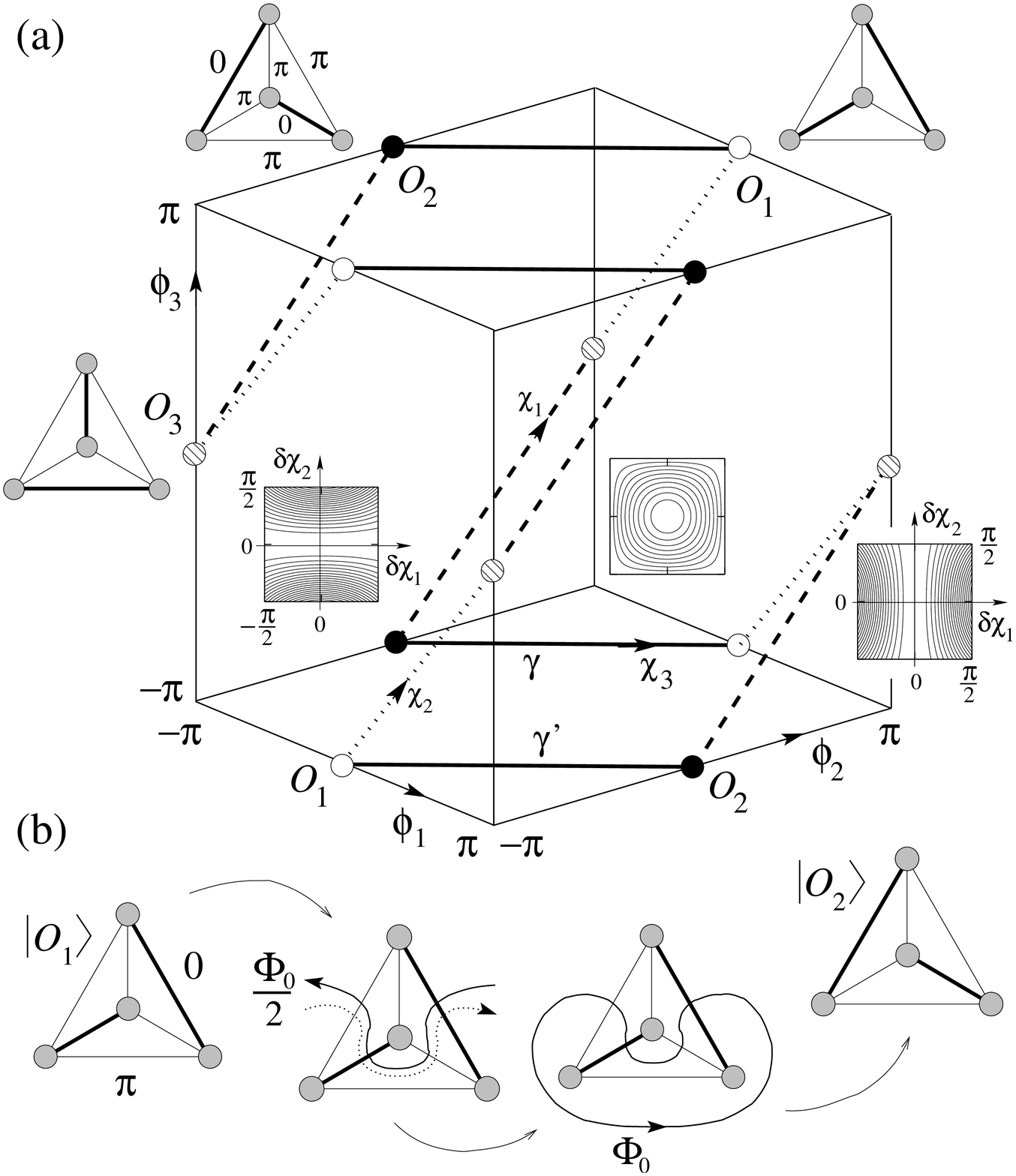}
\caption[]{(a) Continuous degeneracy of the classical minimal energy
  states. We identify 3 highly symmetric minimal states $O_i$,
  $i=1,2,3$ (white, black, and grey dots) involving two opposite
  junctions with a phase difference $\phi_{ij}=0$ and four remaining
  junctions with $\phi_{ij}=\pi$. Three families of lines (solid,
  dashed, dotted) connect these states around the cube $[-\pi,\pi]^3$
  where the energy is continuously degenerate with a value $V_\pi =
  -2 E_J$. The (non-orthogonal) coordinates $\chi_i$ are directed
  along the minima, e.g., $\chi_1=0$, $\chi_2=-\pi$, $\chi_3 \in
  [0,\pi]$ along the reference line $\gamma$ along which the
  change in the potential $V_\pi$ is shown for the values
  $\chi_3 = 0,\pi/2,\pi$. Quantum fluctuations induce potential
  barriers along the minimal lines, separating the classical states
  $O_i$ and transforming them into semi-classical states
  $|O_i\rangle$. (b) Tunneling between the states $|O_i\rangle$
  establishes the low-energy spectrum of the tetrahedron.
  The tunneling process connecting the states $|O_1\rangle$ and
  $|O_2\rangle$ involves two trajectories $\gamma$ and $\gamma'$
  where a fluxon $\Phi_0/2$ cuts through 4 junctions, flipping
  all of them by $\pi$. The phase difference between the two
  trajectories (the solid line corresponds to $\gamma$)
  produces the Aharonov-Bohm-Casher phase obtained when
  taking the fluxon $\Phi_0$ around the islands `1' and `2'.}
  \label{fig:cube}
\end{figure}

We account for the quantum dynamics in the array via the
capacitive term
\begin{equation}
   T = \frac{\hbar^2}{4e^2}\left[
       \sum_{i<j}\frac{C_J}2(\dot\phi_{ij})^2
      + \sum_{i=1}^3\frac{C_g}2 (\dot\phi_i)^2
      + \frac{C_0}2 (\dot\phi_0)^2\right]
\label{T}
\end{equation}
in the Lagrangian ${\cal L} = T-V_\pi$; here, $C_J$ denotes the
capacitance of the junctions, while $C_g$ and $C_0$ are the
capacitances to the ground of the islands $i=1,2,3$ and of the
center/ring island. An additional term $\hbar \sum_i q_i
\dot{\phi}_i$ appears when charges $2e q_i$ are induced on the
islands; we will discuss the effect of this topological term
later. For the isolated tetrahedron of Fig.\ 1(a) we have $C_0 =
C_g$; going over to center of mass ($4\Phi = \sum_i \phi_i$) and
relative coordinates ($\tilde\phi_i = \phi_i-\Phi$) the difference
variables $\tilde{\phi}_{ij} = \phi_{ij}$ pick up an additional
capacitance $C_0/4$ and the corresponding part of the kinetic
energy can be written in the form $T_{\rm rel} = (\hbar^2/16
E_C)\sum_{i<j} \dot{\phi}_{ij}^2$, with the capacitive energy $E_C
= e^2/2C$ and $C = C_J + C_0/4$. The identical expression for the
kinetic energy is obtained for the inverted tetrahedron in Fig.\
1(b) if we choose a large self-capacitance $C_0/C_J\rightarrow
\infty$ for the ring and a small one $C_g/C_J\rightarrow 0$ for
the other islands, then $E_C = e^2/2C_J$; this limit describes the
inverted tetrahedron connected to the outside world via (large)
superconducting wires fixing the phase $\phi_0=0$. The following
discussion applies to both designs of Fig.\ 1; we first assume
that $E_C \ll E_J$, placing the array into the phase-dominated
regime.

\subsection{Semi-classical Analysis}
Next, we account for quantum fluctuations associated with the
degenerate line-minima. It is convenient to introduce the
(non-orthogonal) variables $\chi_k=(\phi_i+\phi_j- \phi_k)/2$ with
$i,j,k \in \{1,2,3\}$ and mutually different, where both the
kinetic and potential energy terms aquire a simpler form, $T_{\rm
rel} = (\hbar^2/4E_c)[\dot{\chi}_1^2+\dot{\chi}_2^2+
\dot{\chi}_3^2]$ and $V_\pi = 2E_J[\cos\chi_1 \cos\chi_2 +
\cos\chi_1 \cos\chi_3 + \cos\chi_2 \cos\chi_3]$. The new
coordinates are directed along the potential minima which are
parametrized by fixing two coordinates to 0 and $\pm\pi$ and have
the third run through the interval $[0,\mp\pi]$; e.g., the
reference line (denoted by $\gamma$) connecting $O_2$ with $O_1$
in Fig.\ 2(a) is parametrized by $\chi_1=0$, $\chi_2=-\pi$,
$\chi_3 \in [0,\pi]$. In the vicinity of $\gamma$ the potential
energy takes the form $V_\pi \approx E_J[-2 + \delta\chi_1^2
(1-\cos\chi_3) + \delta\chi_2^2(1+\cos\chi_3)]$; the two fast
oscillatory modes $\chi_1$ and $\chi_2$ appear with a curvature
which depends on the adiabatic coordinate $\chi=\chi_3$, see Fig.\
2(a). Their zero-point fluctuations produce an induced potential
\begin{equation}
   V_{\rm f}(\chi)=\frac{1}{2}[\hbar\omega_1(\chi)
                        +\hbar \omega_2(\chi)],
   \label{E0}
\end{equation}
with the frequencies of the fast modes
\begin{equation}
   \omega_{1,2}(\chi)
   = \omega_{\rm f} \sqrt{[1 \pm \cos(\chi)]/2}
   \label{omega}
\end{equation}
and the frequency scale $\omega_{\rm f}=\sqrt{8 E_J E_C} /\hbar$.
Near the edges $\chi = 0,\pi$ one of the modes goes to zero, as is
clear from the potential shape shown in Fig.\ 2(a), and we have to
refine our analysis. We then remain with only one fast and two
slow modes. Expanding $V_\pi$ around the point $O_2 = (0,-\pi,0)$
(in $\chi_i$ coordinates) we arrive at the potential $V_\pi =
(E_J/2)[-4+4\delta\chi_2^2 + \delta\chi_1^2\delta\chi_3^2 -
\delta\chi_2^2(\delta\chi_1^2+ \delta\chi_3^2)]$; integration over
the fast mode $\chi_2$ and transformation to momenta provides us
with the Hamiltonian
\begin{equation}
   {\cal H}_{\rm f} \approx -E_C
   \left[\frac{d^2}{d\chi_1^2} + \frac{d^2}{d\chi_3^2}\right]
   +E_J\frac{\delta\chi_1^2\delta\chi_3^2-\kappa(\delta\chi_1^2
   +\delta\chi_3^2)}{2},
   \label{H}
\end{equation}
where $\kappa = \langle\delta\chi_2^2\rangle = (E_C/8E_J)^{1/2}$.
Dimensional analysis tells that the low lying levels of this
quartic anisotropic oscillator are of the order of
\begin{equation}
   \Omega \equiv \omega_{\rm f}
   \left(\frac{E_C}{E_J}\right)^{1/6} \ll \omega_{\rm f};
   \label{Omega}
\end{equation}
the numerical factors determining the exact positions of the
groundstate and of the non-equidistant higher levels have to be
determined numerically and the results are summarized in Table
\ref{O}; the groundstate energy is accurately described by the
expression $\Omega_0/\Omega \approx 0.311-0.129 (E_C/E_J)^{1/6}$,
where the scale dependence of the correction easily follows from
first order perturbation theory in the term $-E_J \kappa
(\delta\chi_1^2+\delta\chi_3^2)/2$.

Repeating this analysis for the other classical line-minima,
we arrive at three distinct quantum states $|O_i\rangle$
(at energies $-2E_J +\hbar \omega_{\rm f}/2 +\hbar\Omega_0$)
associated with the three classical zero-current states $O_i$
described above. These isolated quantum states are generated
through a fluctuation-induced potential reminding about the
Casimir force between metallic plates \cite{casimir_48} or
the van der Waals interaction between neutral atoms \cite{langbein_74}.
This is just the mechanism producing order from disorder
originally proposed by Villain \cite{villain_80,tsvelik_95}
where the huge classical ground state degeneracy (which
does not follow from the symmetry properties of the system)
is removed by quantum fluctuations, here, selecting the three
points $|O_i\rangle$ as new ground states.

The low-energy spectrum near the points $O_i$ exhibits
non-equidistant levels $\Omega_i$ even deep in the semiclassical
regime, allowing for the use of the tetrahedral structure as a
simple Josephson junction qubit of the type introduced in Refs.\
\onlinecite{yu_02,martinis_02}. Moreover, both symmetry arguments
and the numerical data tell that the first excited level is twofold
degenerate, such that we effectively deal with a spin 1 system
with an easy-plane anisotropy $H_{S=1} = (\Omega_1 -\Omega_0) S_z^2$.
\begin{table}
   \caption{\label{O} Ground- and excited states energies near
   the minima $O_i$. The first excited state at $\Omega_1$ is
   doubly degenerate. The parameter $\nu=\nu_\#(E_J/E_C)^{1/3}$
   quantifies the susceptibility of the semi-classical
   ground states $|O_i\rangle$ to applied fluxes, cf.\ (\ref{v2}).}
   \vskip 0.1truecm
   \begin{tabular}{ c r r r r}
   \noalign{\vskip 3 pt}\hline\noalign{\vskip 1.5 pt}
   \hline\noalign{\vskip 3 pt}
   $E_J/E_C$ & $~~~~~\Omega_0/\Omega$ &
                   $~~~~\Omega_1/\Omega$ &
                   $~~~~\Omega_2/\Omega$ & $~~~~~~\nu_\#$\\
   \noalign{\vskip 3 pt}\hline\noalign{\vskip 3 pt}
   $\infty$ & 0.311 & 0.70 & 0.99 & 1.00\\
   \noalign{\vskip 5 pt}
   1000 & 0.271 & 0.57 & 0.69 & 1.23\\
   \noalign{\vskip 5 pt}
   100  & 0.250 & 0.51 & 0.58 & 1.53\\
   \noalign{\vskip 5 pt}
   10   & 0.225 & 0.40 & 0.43 & 2.50\\
   \noalign{\vskip 3 pt}\hline\noalign{\vskip 1.5 pt}
   \hline\noalign{\vskip 3 pt}
\end{tabular}
\end{table}

Before going to the full quantum description, let us discuss the
above semiclassical version of the device, as it exhibits a number
of interesting features by itself. First, the potential (\ref{E0})
defines a doubly-periodic junction \cite{blatter_01} with two
distinct minima. The potential $V_{\rm f}(\chi)$ can be mapped out
experimentally through the measurement of the Josephson current
$I_J(\chi)$ that can be pushed through the structure. E.g., fixing
the phase $\phi_2$ between the central island and the island `2'
via a flux-biased external loop defines the two classical minimal
solutions $(\phi_2 -\pi, 0, \pi)$ and $(0,-\pi,\phi_2)$ (in
$\chi_i$-coordinates). The running coordinate $\chi = \chi_1$ or 
$\chi = \chi_3$ is equal to $\phi_2$ (up to a trivial shift) 
and thus related to the external bias flux $\Phi$ via $\chi =
-2\pi\Phi/\Phi_0$. The current $I=-c\partial_\Phi E$ then is given
by the expression $I_J(\chi) = (2e/\hbar)\partial_\chi V_{\rm f}(\chi)$
and is double periodic in the interval $0 \leq \Phi < \Phi_0$ (we
define the charge of the electron as $-e$ and $e >0$).
Alternatively, one may measure the frequencies $\omega_{1,2}(\chi)$, 
cf.\ Eq.\ (\ref{omega}), directly via the resonant absorption 
of an $ac$-signal.

At nonzero temperatures (but still $T \ll E_J$) the induced
potential is driven thermally and involves the free energy of the
two fast oscillating modes,
\begin{equation}
   F_{\rm f} (\chi,T)  = \sum_{i=1}^2 \left[\frac{\hbar\omega_i(\chi)}{2}
   + T\log\left(1-e^{-\hbar\omega_i(\chi)/T}\right) \right].
   \label{Find}
\end{equation}
Thermal fluctuations become relevant for temperatures $T > \hbar
\omega_{\rm f}$ and lead to an increase in the barrier $\delta
F_{\rm f}(T) \equiv F_{\rm f}(\pi/2,T)-F_{\rm f}(0,T)$,
\begin{eqnarray}
   \delta F_{\rm f} (T) &=& (\hbar\omega_{\rm f}/2)
   [\sqrt{2}-(1+2\Omega_0/\omega_{\rm f})]
   \nonumber \\
   &+&T \log[\omega_{\rm f}/2\Omega_0].
   \label{dFind}
\end{eqnarray}
As a result, rather then decreasing, the phase stiffness in the
tetrahedron increases with temperature and hence the Josephson
current $I_J(T) \propto \partial_\Phi F_{\rm f}(\Phi,T)$ increases
with temperature until the fluctuation-induced potential
disappears due to level broadening: thermally induced
quasiparticles produce a level broadening $\hbar/RC \sim (E_J
E_C/\hbar\Delta) \exp(-\Delta/T)$, where we have used the
Ambegaokar-Baratoff relation $I_c R \sim \Delta/e$. This
broadening should remain small on the scale of the level
separation $\hbar \omega_{\rm f}$, from which we obtain the
condition that $T < \Delta/\ln(\hbar\omega_{\rm f}/\Delta)$.
Beyond this temperature, the Josephson current is expected to
decrease again, resulting in a non-monotonic behavior of $I_J(T)$.

In order to arrive at a fully quantum mechanical description of
the tetrahedron, we have to account for the tunneling processes
between the points $O_i$, cf.\ Fig.\ 2. It is important to note
that each set of 4 mid-edge points residing in one plane $\phi_i =
0$ has to be identified with one quantum mechanical state
$|O_i\rangle$; on the other hand, the pair of classically
degenerate lines connecting two states $|O_i\rangle$ and
$|O_j\rangle$ in one of the faces describe different tunneling
trajectories, which have to be added coherently in order to arrive
at the tunneling matrix element between the two states (the other
two trajectories on the opposite face are equivalent). Let us
concentrate on the tunneling process between $|O_1\rangle$ and
$|O_2\rangle$; the two tunneling trajectories follow the lines
parametrized by $\chi_1,\chi_2 \in \{0,-\pi\}$, $\chi_3 \in
[0,\pi]$, see Fig.\ 2(a). The tunneling processes described by
these two trajectories flip the phase across the four junctions
`31', `01', `02', and `23' by $\pm \pi$, which corresponds to a
fluxon $\Phi_0/2$ traversing the tetrahedron as shown in Fig.\
2(b) (the same arguments apply to the connected tetrahedron). The
phase difference between the two trajectories then corresponds to
taking a full fluxon $\Phi_0$ around the two islands `1' and `2',
which translates into the Aharonov-Bohm-Casher phase $\exp[2\pi i
(q_1 +q_2)]$, with $q_i$ the charge on the island `$i$' measured
in units of $2e$ (see Ref.\ \onlinecite{ivanov_02} for a detailed
discussion of charge-induced interference effects in small
superconducting structures; these phases are generated by the
topological term $\hbar \sum_i q_i
\dot{\phi}_i$ in the Lagrangian ${\cal L}$ in the presence of
charges $2e q_i$, cf.\ the note below (\ref{T})). Combining this
phase factor with the modulus $|a|$ we arrive at the tunneling
amplitude
\begin{equation}
   t_{12} = -2|a|\cos[\pi(q_1+q_2)]
   \label{tij}
\end{equation}
between the states $|O_1\rangle$ and $|O_2\rangle$; a similar
analysis provides the amplitudes for all the other pairs. In the
absence of any charge frustration (i.e., for integer charge $q_i$
on each island) the system gains energy from hopping and hence $t
< 0$, thus defining the sign in (\ref{tij}). The modulus $|a|$ of
the tunnelling amplitude follows from the semi-classical
description of the one-dimensional motion under the barrier
$V_{\rm f}(\chi)$ as given by Eq.\ (\ref{E0}) and takes the form
$|a| \approx [\hbar/T(\Omega_0)] \exp[-S_{\rm f}(\Omega_0)]$, with
$T(\Omega_0)$ the classical period of motion and $S_{\rm
f}(\Omega_0)$ the dimensionless action \cite{LL}, both evaluated
at the ground state energy $\hbar \Omega_0$,
\begin{eqnarray}
   && S_{\rm f} = \left(\frac{32 E_J}{E_C}\right)^{\!\! 1/4}
   \!\!\int_0^{\chi_0} \!\!\!
   d\chi\, [\sqrt{2}\cos(\chi/2)-1-
   2\Omega_0/\omega_{\rm f}]^{1/2}
   \nonumber \\
   &&\quad \sim
   1.88 \left(\frac{E_J}{E_C}\right)^{1/4},\qquad (E_J/E_C)^{1/4} \gg 1,
   \label{action}\\
   &&\quad T = \frac{8\hbar}{E_J}
   \left(\frac{\sqrt{2}\Omega_0}{\Omega}\right)^{1/2}
   \left(\frac{E_J}{E_C}\right)^{2/3},
   \label{period}
\end{eqnarray}
where $\chi_0 = 2 \arccos[(1+2\Omega_0/\omega_{\rm f})/\sqrt{2}]$.
The semi-classical analysis describing tunneling at the correct
ground state energy $\hbar\Omega_0$ gives very accurate results,
see below; the simple asymptotic form (\ref{action}) describes
tunneling from the
bottom of the well and is valid only deep in the quasi-classical
regime. The result obtained here for the tetrahedron is smaller
than the usual tunneling action $S \propto (E_J/E_C)^{1/2}$ of a
Josephson junction device involving only the square root of the
parameter $E_J/E_C$; the unconventional dependence on the ratio
$E_J/E_C$ is a consequence of the fluctuation-induced (rather than
classical) barrier and puts less stringent requirements (e.g.,
with respect to the smallness of the junction) on the fabrication
process of this new type of qubits.

The above tunneling action can be probed via a measurement of the
current-voltage characteristic of the device: applying a fixed
bias current across two islands, a finite voltage appears due to
quantum and thermally induced phase slips. At low temperatures the
appearance of quantum phase slips involves the finite action
$S_{\rm ps} = 2\hbar S_{\rm f}$ and results in an exponentially
small resistivity $R \propto \exp(-2 S_{\rm f})$. At higher
temperatures $T > U_{\rm ps} \hbar /S_{\rm ps}$ the phase slips
are created thermally via activation over the barrier $U_{\rm ps}
= \delta F_{\rm f}$. The phase-slip induced resistivity then
exhibits an unconventional saturation at high temperatures:
increasing $T$ beyond $\hbar \omega_{\rm f} (E_C/ E_J)^{1/4}$, the
resistivity first increases with temperature. As $T >\hbar
\omega_{\rm f}$ the barrier $\delta F_{\rm f}$ itself increases
linearly in $T$ and the further rise of $R$ {\it saturates} at a value
$R \propto \Omega_0/\omega_{\rm f}$; hence care has to be taken
not to confuse the saturation in $R$ at high temperatures with the
(exponentially small) quantum resistance surviving as $T
\rightarrow 0$.

Let us return to the quantum description of our tetrahedron and
study its low-energy spectrum as determined by the quantum
coherent oscillations between semi-classical states $|O_i\rangle$;
the appearance of the island charges $2e\,q_i$ in the tunneling
amplitudes $t_{ij}$ manifests itself in this level structure.
Assuming a uniform distribution of the total charge $2e\,q$ on the
isolated tetrahedron, the matrix elements take the form $t_{ij} =
t = -2 |a| \cos(\pi q/2)$ and we have to diagonalize the matrix
\begin{equation}
   H_{O} = \left(
         \begin{array}{lcr}
         0 &t& t \\
         t &0 &t \\
         t & t& 0
         \end{array}
         \right).
   \label{Ht}
\end{equation}
Depending on the value of $q$, the low-energy spectrum of the
isolated tetrahedron splits into singlets and doublets involving
the energies $E_{\rm s} = 2t$ and $E_{\rm d} = -t$, or remains
tri-fold degenerate with $E_{\rm t}(t=0) = 0$ (here, energies are
measured with respect to the unperturbed value $-2E_J+\hbar
(\omega_{\rm f}/2+\Omega_0)$); the ground state is a
\begin{equation}
   \begin{array}{lrr}
   {\rm singlet} \,\, {\rm if} & q=4k, &   \quad {\rm even}, \\
   {\rm doublet} \,\, {\rm if} & q=4k+2, & \quad {\rm even},\\
   {\rm triplet} \,\, {\rm if} & q=2k+1, & \quad {\rm odd}.
   \end{array}
   \label{triplet}
\end{equation}
In the connected tetrahedron of Fig.\ 1(b) the charge is not
quantized on the inner islands; however, the dependence
(\ref{tij}) of the tunneling amplitude on the island charges $q_i$
remains valid and the above spectrum is recovered under
appropriate biasing of the inner islands with either zero,
one-quarter, or one-half Cooper-pair. Biasing the tetrahedron into
the charge state $q_i=1/2$ then establishes a ground state doublet
with eigenstates
\begin{eqnarray}
   |+\rangle &=&
   [|O_1\rangle+\zeta|O_2\rangle+\zeta^*|O_3\rangle]/\sqrt{3},
   \nonumber\\
   |-\rangle &=&
   [|O_1\rangle+\zeta^*|O_2\rangle+\zeta|O_3\rangle]/\sqrt{3},
   \label{ev}
\end{eqnarray}
suitable for the implementation of a quantum bit; here, $\zeta =
\exp(2\pi i/3)$. These degenerate states (\ref{ev}) involve bonds
resonating with opposite chirality in the device, see Fig.\ 2(a),
and are reminiscent of the resonating dimer bonds in the
topologically protected qubits discussed in Refs.\
\onlinecite{ioffe1_02,ioffe2_02}. Also, such chiral states appear as
degenerate spin-singlet ground states in the vanadium tetrahedron
of the pyrochlore system \cite{tsunetsugu_02}. The doublet 
$|\pm\rangle$ is protected by the gap $\Delta_{\rm d} = 3 t$,
separating it from the next excited (singlet) state
\begin{equation}
   |0\rangle =
   [|O_1\rangle+|O_2\rangle+|O_3\rangle]/\sqrt{3}.
   \label{0}
\end{equation}
Combining the results (\ref{tij}), (\ref{action}), and
(\ref{period}) we obtain the protective and operational energy
scale $t$ of the qubit,
\begin{equation}
   \frac{t}{E_J} = \frac{1}{4}
   \left(\frac{\Omega}{\sqrt{2}\Omega_0}\right)^{1/2}
   \left(\frac{E_C}{E_J}\right)^{2/3}
   \exp[-S_{\rm f}(\Omega_0)].
   \label{Delta_d}
\end{equation}

\subsection{Charge limit}
We briefly extend our analysis to the charge-dominated regime
with $E_C \gg E_J$. It is convenient to go over to a Hamiltonian
description; starting from the above Lagrangian (cf.\ Eqs.\
(\ref{EJ}) and (\ref{T})) and eliminating the variable $\phi_0$
one obtains the expression
\begin{eqnarray}
   H&=& E_C[\bar{Q}_1^2+\bar{Q}_2^2+\bar{Q}_3^2+(\bar{Q}_1
           +\bar{Q}_2+\bar{Q}_3)^2]
   \label{Ham2} \\
   &&\quad
   + V_\pi(\phi_1,\phi_2,\phi_3),
   \nonumber
\end{eqnarray}
where $\bar{Q}_i = -i\partial_{\phi_i}-q_i$ and $q_i$ is the
induced charge on the $i$-th island (in units of $2e$), $q_i =
\sum_j C_{ij} V_j$ for the connected tetrahedron, while an
additional term $[q-(C/4)\sum_j V_j]/4$ with $C = \sum_{ij}
C_{ij}$ the total capacitance, has to be added for the isolated
tetrahedron \cite{ivanov_02} (here, $C_{ij}$ denotes the
capacitance matrix, see (\ref{T}), and $V_i$ are the bias
potentials applied to the islands). For the isolated tetrahedron
the total charge $q=\sum_0^3 q_i$ is integer, while the total
induced charge $q=\sum_1^3 q_i$ can take any value in the
connected device. The Hamiltonian (\ref{Ham2}) describing both
devices becomes identical under symmetric bias and for specific
values $q_i = k/4$ with $k$ an integer; under these conditions the
maximal symmetry $S_4$, i.e., the tetrahedral symmetry $T_d$, is
established. Note that a symmetric bias with equal charges $q_i =
q/3$ on the three inner islands of the connected tetrahedron in
general guarantees only for a $S_3$ symmetry.

We determine the spectrum for the uniformly charged isolated
tetrahedron with $q=4k+2$. In the limit $E_C \gg E_J$ the
operators $Q_i$ take on integer values and the charging term is
minimized by distributing two bosons onto the four sites avoiding
double occupancy (in this limit, the term $E_C
(\bar{Q}_1+\bar{Q}_2+\bar{Q}_3)^2$ in (\ref{Ham2}) describes the
charging energy $E_C \bar{Q}_0^2$ of the middle island). The
resulting six states `$01$', `$02$', `$03$', `$23$', `$31$', and
`$12$' are degenerate with an energy $E_0 = 2E_C$. The hopping
term $V_\pi$ lifts this degeneracy through the mixing via the
Josephson coupling $E_J$: each state `$ij$' hosting Cooper-pairs
on the islands `$i$' and `$j$' exchanges particles with all other
states except for `$kl$', where $k,l \neq i,j$. The Hamiltonian
describing the mixing of the six two-Boson states may be written
as a matrix product
\begin{equation}
   H_{\rm\scriptscriptstyle 2B} = \left(
         \begin{array}{lcr}
         0 &\tilde{t}& \tilde{t} \\
         \tilde{t} &0 &\tilde{t} \\
         \tilde{t} & \tilde{t}& 0
         \end{array}
         \right) \otimes
   \left(
     \begin{array}{lr}
      1 & 1 \\
      1 & 1
       \end{array}
   \right),
   \label{H_2B}
\end{equation}
where the sign of the tunneling amplitude $\tilde{t} = E_J/2>0$ is
positive for our frustrated tetrahedron (again, we choose a
symmetric gauge with all Josephson couplings reversed in sign).
The eigenvalues of the direct matrix product in (\ref{H_2B}) are
given by the product of the eigenvalues $(2\tilde{t}, -\tilde{t},
-\tilde{t})$ of its first factor and those of the second factor,
$(0, 2)$; correspondingly we find the first 6 levels at the
energies
\begin{equation}
   \begin{array}{lrr}
   E_{\rm d} = - E_J & \quad {\rm doublet}, \\
   E_{\rm t} = 0     & \quad {\rm triplet},  \\
   E_{\rm s} = 2 E_J & \quad {\rm singlet}.
   \end{array}
   \label{levels}
\end{equation}
The first excitation now is a triplet rather than the singlet
state found in the opposite limit $E_J \gg E_C$; hence, decreasing
$E_J/E_C$ from large values, the singlet and triplet energies
cross each other and the first excitation gap changes over from
$3t$ (at $E_J \gg E_C$) to $2\tilde{t}= E_J$ (for $E_C\gg E_J$).
The precise location where this crossing appears can be found from
the numerical analysis described below. The same analysis can be
repeated for the connected tetrahedron; the six states degenerate
under the capacitive term then involve either one or two charges
on the three inner islands and the mixing term $V_\pi$ describes
charges hopping between the three islands as well as hopping of
one charge to and from the ring.

\subsection{Numerical Results}
The above results can be verified numerically via diagonalization
of the Hamiltonian (\ref{Ham2}) with the help of a Lanczos
algorithm; in the charge basis the mixing term $V_\pi$ then
describes the hopping of charges between the islands. Going to a
phase representation, the bias charges $q_i$ are conveniently
accounted for via the boundary condition for the wave function,
$\Psi_n(\phi_1+2\pi \delta_{1k}, \phi_2+2\pi\delta_{2k},
\phi_3+2\pi\delta_{3k})= \exp(-2\pi i q_k) \Psi_n(\phi_1, \phi_2,
\phi_3)$, $k=1,2,3$, after a suitable gauge transformation
\cite{ivanov_02}. The results of such an analysis for the charge
state $q_i=1/2$ is shown in Fig.\ 3, where the excitation gap
$\Delta_{\rm d}$ protecting the qubit states against higher
excitations is shown as a function of $E_J/E_C$. The crossover
from the charge to the phase dominated regime, where the singlet
and triplet excited states cross, takes place at $E_J/E_C \approx
5$. The analytic results (\ref{Delta_d}) and (\ref{levels})
describe well the data away from the crossover regime. One expects
the quasi-classic result to become exact in the limit of large
$E_J/E_C$; however, the result (\ref{Delta_d}) has been calculated
using the one-dimensional approximation $V_{\rm f}$ for the 
potential and one cannot expect perfect agreement with the
numerical data. Still, the quasi-classic approximation turns 
out accurate over a very large regime extending down to 
parameters $E_J/E_C$ of order 10: scaling the dashed line 
in Fig.\ 3 by 0.8 the quasi-classic result cannot be 
distinguished from the numerical data for $E_J/E_C > 20$. 
\begin{figure}[ht]
\includegraphics[scale=0.4]{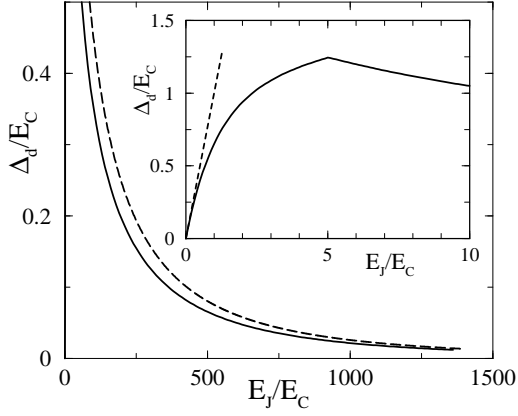}
\caption[]{Excitation gap $\Delta_{\rm d}$ protecting the qubit
  doublet against higher excitations as a function of $E_J/E_C$; the
  dashed line is the semi-classic result based on (\ref{Delta_d}). The
  inset shows an expanded view illustrating the crossing of the
  singlet and triplet excited states as the system enters the charge
  dominated regime at low values of $E_J/E_C$;
  the dashed line is the analytic result (\ref{levels}) valid
  in the charge dominated regime $E_C \gg E_J$.
  Data obtained with help of a Lanczos algorithm operating in the
  charge representation with 27 charge states between $q=\pm 13$.}
  \label{fig:tetrahedron_3}
\end{figure}

The numerical results show that the suppression of the tunneling
amplitude $t$ is indeed small in the tetrahedron. E.g., choosing a
value $E_J/E_C$ of order 100 the energy scale $t$ of the qubit is
suppressed by a factor $\sim 2/1000$ with respect to the energy
scale $E_J$ of the junctions. For a conventional device this
suppression involves an action $S \approx c\sqrt{E_J/E_C}$, with
the numerical $c$ depending on the specific setup. E.g., for the
4-junction loops studied in Ref.\ \onlinecite{blatter_01} the
numerical $c \approx 1.6$ and choosing the same value of $E_J/E_C$
this implies a suppression of quantum fluctuations by a factor
$\sim \exp(-16)\sim 10^{-7}$; such a device then resides deep in
the semi-classical regime and quantum effects are heavily
suppressed.

In summary, we can tune our tetrahedral structure such as to
realize a doubly-degenerate groundstate corresponding to a
spin-1/2 system in zero magnetic field; the device can be realized
using moderately large junctions with $E_J/E_C$ of order 100 while
keeping an appreciable operational energy scale $t$, a consequence
of the particular frustration in the device. Below, we will show
that this ground state remains robust to quadratic order in the
external noise and hence provides a suitable starting point for
the construction of a qubit; we will concentrate on this
doubly-degenerate case and its use for quantum computing in the
following.

\subsection{External fields}
Fabrication errors and external bias induce splittings and shifts
in the levels. With respect to the qubit's functionality, random
external signals produce decoherence, while prescribed bias
signals are used for its manipulation. Fabrication errors mainly
affect the coupling $E_J$ of the junctions --- we denote the
corresponding deviations by $\epsilon_{ij} E_J$ (note that the
fabrication tolerance is improved for larger junctions, leading to
smaller values of $\epsilon_{ij}$). External bias signals appear
randomly through fluctuations in the local magnetic field and
through stray charges; artificially generated signals can be
applied through properly placed current loops biasing the
sub-loops of the tetrahedron (currents $i_i$ in Fig.\
\ref{fig:tetrahedron_1}) and through capacitive charging of the
islands (via voltages $v_i$). We denote the corresponding bias
fields by $\delta^{\scriptscriptstyle \Phi}_i\equiv -2\pi
\Phi_i/\Phi_0$ and $\delta^{\scriptscriptstyle Q}_i \equiv \delta
q_i$, $i=1,2,3$. E.g., for the isolated tetrahedron the flux
$\Phi_3$ penetrating the sub-loop `0-1-2' is encoded in a bias
angle $\delta^{\scriptscriptstyle \Phi}_{3} \equiv
\delta(\phi_2-\phi_1)= -2\pi\, \Phi_3/\Phi_0$ along the link
`1-2'; equivalent definitions apply to the other sub-loops with
the cyclic replacement $3 \rightarrow 1 \rightarrow 2 \rightarrow
3$. For the connected tetrahedron we have to account for the total
flux $\Phi$ threading the outer ring; assuming a symmetric setup,
this flux will induce the phase shifts $\delta = -2\pi\Phi /3
\Phi_0$ in each of the segments `m1-m2', `m2-m3', and `m3-m1',
cf.\ Fig.\ 1(b). The bias angles $\delta^{\scriptscriptstyle
\Phi}_i$ along the links `$j$-$k$' then have to account for this
flux via the modified form $\delta^{\scriptscriptstyle \Phi}_{i} =
2\pi\, (\Phi_i-\Phi/3) /\Phi_0$, where $\Phi_i$ denotes the flux
through the loop `m$j$-m$k$-$k$-$j$-m$j$'. Note that we have
assumed that the currents circulating in the outer ring are still
sufficiently small such as not to produce significant self-fields.
While the charge bias $\delta^{\scriptscriptstyle Q}_i$ only
affects the tunneling matrix elements $t_{ij}$, i.e., the kinetic
energy term, the flux bias $\delta^{\scriptscriptstyle \Phi}_{i}$
modifies both the potential and the kinetic energy terms in the
Lagrangian.

We first determine the modification of the potential energy
$V_\pi$, cf.\ (\ref{EJ}), due to applied fluxes
$\delta^{\scriptscriptstyle \Phi}_i$,
\begin{eqnarray}
   &&\frac{V_\pi}{2E_J}
   = \cos\!\Big(\chi_1+\frac{\delta^{\scriptscriptstyle \Phi}_3}{2}\Big)
         \cos\!\Big(\chi_2-\frac{\delta^{\scriptscriptstyle \Phi}_3}{2}\Big)
         +\cos\!\Big(\chi_3+\frac{\delta^{\scriptscriptstyle \Phi}_2}{2}\Big)
        \nonumber \\
        && \times
        \cos\!\Big(\chi_1-\frac{\delta^{\scriptscriptstyle \Phi}_2}{2}\Big)
        +\cos\!\Big(\chi_2+\frac{\delta^{\scriptscriptstyle \Phi}_1}{2}\Big)
         \cos\!\Big(\chi_3-\frac{\delta^{\scriptscriptstyle \Phi}_1}{2}\Big).
         \label{dV}
\end{eqnarray}
Assuming small perturbations, we expand (\ref{dV}) in
$\delta^{\scriptscriptstyle \Phi}_i \ll \omega_{\rm f}/E_J$; we
concentrate on the point $O_2 =(0,-\pi,0)$ and combine the result
with the kinetic term to arrive at the Hamiltonian (\ref{H}) with
the additional term
\begin{eqnarray}
   \frac{\delta H}{E_J}
   &=& -
   \Big[\delta^{\scriptscriptstyle \Phi}_1(\delta\chi_3-\delta\chi_2)
   +\delta^{\scriptscriptstyle \Phi}_2(\delta\chi_3-\delta\chi_1)
   \nonumber \\
   &+&\delta^{\scriptscriptstyle \Phi}_3 (\delta\chi_2-\delta\chi_1)
   -\frac{1}{2}({\delta^{\scriptscriptstyle \Phi}_1}^2
   +{\delta^{\scriptscriptstyle \Phi}_3}^2
   -{\delta^{\scriptscriptstyle \Phi}_2}^2)\Big].
   \label{dH}
\end{eqnarray}
Classically, the force term in (\ref{dH}) lowers the energy
indefinitely as the system runs away along the degenerate
classical minimal lines; e.g., a perturbation
$\delta^{\scriptscriptstyle \Phi}_2>0$ produces a runaway either
along the $(0,0,\delta\chi_3)$ direction or along the
$(-\delta\chi_1,0,0)$ direction. However, quantum fluctuations
generate a finite potential along these lines, resulting in a
linear response in the coordinates $\delta \chi_i$ and a quadratic
change in energy $v_2$. Indeed, second-order perturbation theory
in the force term of (\ref{dV}) produces the result
\begin{eqnarray}
   \frac{v_2}{E_J}
   &=& - \nu
   \Big[(\delta^{\scriptscriptstyle \Phi}_1
   +\delta^{\scriptscriptstyle \Phi}_2)^2
   +(\delta^{\scriptscriptstyle \Phi}_2
   +\delta^{\scriptscriptstyle\Phi}_3)^2\Big]
   \nonumber \\
   &+&(\epsilon_{02}-\epsilon_{03}-\epsilon_{01})
   +(\epsilon_{31}-\epsilon_{12}-\epsilon_{23}),
   \label{v2}
\end{eqnarray}
with
\begin{equation}
   \nu \approx 1.0\, \left(E_J/E_C\right)^{1/3}
   \label{nu}
\end{equation}
obtained from a numerical solution of the perturbed
eigenvalue problem combining (\ref{H}) and (\ref{dH}). In
(\ref{v2}) we have dropped the term $(E_J/2)
({\delta^{\scriptscriptstyle \Phi}_1}^2
+{\delta^{\scriptscriptstyle \Phi}_3}^2
-{\delta^{\scriptscriptstyle \Phi}_2}^2)$ as it is small by a
factor $(E_C/E_J)^{1/3}$ compared to the leading term; also, we
have added a term due to deviations $\epsilon_{ij}E_J$ in the
junction couplings. Equivalent expressions for the other minima
follow from cyclic permutation of the indices. As a result, we
obtain the relative shifts $v_{ij} \equiv v_{j} - v_{i}$
($i,j,k=1,2,3$, $k \neq i,j$)
\begin{eqnarray}
   \frac{v_{ij}}{E_J}
   &=&\nu
   (\delta^{\scriptscriptstyle \Phi}_i
   -\delta^{\scriptscriptstyle \Phi}_j)
   (\delta^{\scriptscriptstyle \Phi}_i
   +\delta^{\scriptscriptstyle \Phi}_j +2\delta^{\scriptscriptstyle
   \Phi}_k)
   \label{shift_Vij}\\
   &&\quad +2(\epsilon_{0j}-\epsilon_{0i}+\epsilon_{ki}-\epsilon_{jk})
   \nonumber
\end{eqnarray}
in the minima. First, we note that small random fluxes do not
affect these positions in the first order of these fluctuations;
the corrections appearing in quadratic order then are small.
Second, we note that fabrication errors $\epsilon_{ij}$ in the
junction couplings can be compensated by appropriate choices of
bias fluxes ${\delta^{\scriptscriptstyle \Phi}_i}$.

In the determination of the perturbed tunnneling matrix elements
$t_{ij}$, we ignore the modifications arising due to fabrication
errors and concentrate on the effects of flux and charge signals,
random or externally applied. By way of example, we calculate the
tunneling matrix elements $t_{12}$ and $t_{21}$ connecting the
states $|O_1\rangle$ and $|O_2\rangle$. The presence of perturbing
fluxes $\delta^{\scriptscriptstyle \Phi}_1$ and
$\delta^{\scriptscriptstyle \Phi}_2$ shifts the potential $V_\pi$
by $v$ along the line $\gamma$, to lowest order in
$\delta^{\scriptscriptstyle \Phi}_i$, $v (\chi) = - E_J\, \sin\chi
\,(\delta^{\scriptscriptstyle \Phi}_1 + \delta^{\scriptscriptstyle
\Phi}_2)$. This shift produces the changes $\delta S_{12}^{\pm} =
\mp s (\delta^{\scriptscriptstyle \Phi}_1 +
\delta^{\scriptscriptstyle \Phi}_2)$ (the correction $\delta
S_{12}^+$ applies to the trajectory $\gamma$ in Fig.\ 2(a)) with
\begin{equation}
   s \approx 1.5 \, \left(E_J/E_C\right)^{3/4}
   \label{s}
\end{equation}
in the action $S_{\rm f}$ determining the modulus $|a|$ of the
tunneling amplitude (\ref{tij}); as before, the expression
(\ref{s}) is valid deep in the semi-classical regime.

The presence of perturbing charges $\delta^{\scriptscriptstyle
Q}_1$ and $\delta^{\scriptscriptstyle Q}_2$ modifies the
Aharanov-Bohm-Casher phase associated with the two trajectories (a
charge $Q$ encircling a flux $\Phi$ counter clockwise produces the
phase $\exp[2\pi i (\Phi/\Phi_0)(Q/2e)]$): they pick up the
additional phases $\exp[\pm i\pi( \delta^{\scriptscriptstyle Q}_1
+ \delta^{\scriptscriptstyle Q}_2)]$, with the plus sign belonging
to the trajectory $\gamma$, cf.\ Fig.\ 2(b). Combining the
perturbations in the fluxes and charges, the change in the
tunneling amplitudes $\delta t_{12} = t^s_{12}+i t^a_{12} = \delta
t^*_{21}$ takes the form
\begin{equation}
   \frac{\delta t_{12}}{t} = \frac{
   e^{-\delta S_{12}^{+}
   +i\pi(\delta^{\scriptscriptstyle Q}_1+
   \delta^{\scriptscriptstyle Q}_2)}
   +e^{-\delta S_{12}^{-}
   -i\pi(\delta^{\scriptscriptstyle Q}_1+
   \delta^{\scriptscriptstyle Q}_2)}-2}{2};
   \label{shift_t}
\end{equation}
expanding the exponentials, the symmetric and antisymmetric parts
are given by the expressions
\begin{eqnarray}
   t^s_{12} &=& t \left[
   s^2 (\delta^{\scriptscriptstyle \Phi}_1
   + \delta^{\scriptscriptstyle \Phi}_2)^2/2
   - \pi^2 (\delta^{\scriptscriptstyle Q}_1
   + \delta^{\scriptscriptstyle Q}_2)^2/2\right],
   \nonumber\\
   t^a_{12} &=& \pi \, s t (\delta^{\scriptscriptstyle \Phi}_1
   + \delta^{\scriptscriptstyle \Phi}_2)
    (\delta^{\scriptscriptstyle Q}_1
   + \delta^{\scriptscriptstyle Q}_2);
   \label{tsa}
\end{eqnarray}
further terms quadratic in $\delta^{\scriptscriptstyle \Phi}$
(e.g., arising from the next term in the expansion of $S_{\rm f}$)
are small by the factor $(E_C/E_J)^{1/4}$. With respect to the
qubit's stability, we note that all perturbations appear in second
order of the small quantities $\delta^{\scriptscriptstyle \Phi}_i$
and $\delta^{\scriptscriptstyle Q}_i$. In the further analysis
below we will drop the flux bias term $\propto
{\delta^{\scriptscriptstyle \Phi}_i}^2$ in $t_{ij}^s$ against the
(parametrically large) shifts $v_i$ in the potential energy.

We determine the new energy levels perturbatively: the
perturbation $\delta H$, written in the space of semi-classical
ground states $|O_i\rangle$, takes the form
\begin{equation}
   \delta H_{O} = \left(
         \begin{array}{ccc}
         v_{1} &t^s_{12}+i t^a_{12}& t^s_{31}-i t^a_{31} \\
         t^s_{12}-i t^a_{12} &v_{2} &t^s_{23}+i t^a_{23} \\
         t^s_{31}+i t^a_{31} & t^s_{23}-i t^a_{23}& v_{3}
         \end{array}
         \right).
   \label{dH_O}
\end{equation}
Next, we find the corresponding matrix elements $\langle
\pm |\delta H| \pm \rangle$ in the projected space
spanned by the doublet $|\pm \rangle$, cf.\ (\ref{ev}). It is
convenient to cast the result of this calculation into the form
\begin{equation}
   H_{\rm qubit} = e_{\rm d} \, {\bf 1} + {\bf h}\cdot {\usigma}
   \label{Hspin}
\end{equation}
with $\usigma= (\sigma_x,\sigma_y,\sigma_z)$ the Pauli matrices;
switching of the `magnetic' fields $h_x$ and $h_z$ then produces
the standard amplitude- and phase-shift operations required for
the manipulation of the individual qubit. The shift $e_{\rm d}$
and the effective `magnetic' field ${\bf h}$ are given by the
expressions
\begin{eqnarray}
   e_{\rm d} &=& \frac{1}{3}[v_{1}+v_{2}+v_{3}
             -(t^s_{12}+t^s_{23}+t^s_{31})],\nonumber \\
   h_x &=& \frac{1}{3}\Bigl[v_{1}-\frac{1}{2}(v_{2}+v_{3})
           +(2t^s_{23}-t^s_{12}-t^s_{31})\Bigr],\nonumber \\
   h_y &=& \frac{1}{2\sqrt{3}}(v_{3}-v_{2})
         +\frac{1}{\sqrt{3}}(t^s_{12}-t^s_{31}),\nonumber \\
   h_z &=& - \frac{1}{\sqrt{3}}(t^a_{12}+t^a_{23}+t^a_{31});
   \label{fields}
\end{eqnarray}
note that the perturbations $v_i$ in the potential come with the
large amplitude $E_J$, while those in the kinetic energy
($t^{s,a}$) involve the smaller energy scale $t$ of the tunneling
matrix element; in (\ref{fields}) we keep both terms as we might
choose to manipulate the qubit via changes in the charges
$\delta^{\scriptscriptstyle Q}$ alone. Expressing the
perturbations in terms of the flux- and charge-parameters
$\delta^{\scriptscriptstyle \Phi}_i$ and
$\delta^{\scriptscriptstyle Q}_i$ we obtain the results
\begin{eqnarray}
            e_{\rm d} &=&
            -\frac{4\nu E_J}{3}\,
            ({\delta^{\scriptscriptstyle \Phi}_1}^2
            +{\delta^{\scriptscriptstyle \Phi}_2}^2
            +{\delta^{\scriptscriptstyle \Phi}_3}^2
            +\delta^{\scriptscriptstyle \Phi}_1
            \delta^{\scriptscriptstyle \Phi}_2
            +\delta^{\scriptscriptstyle \Phi}_1
            \delta^{\scriptscriptstyle \Phi}_3
            +\delta^{\scriptscriptstyle \Phi}_2
            \delta^{\scriptscriptstyle \Phi}_3)
            \nonumber \\ &&
            +\frac{t\pi^2}{6}\,
            [(\delta^{\scriptscriptstyle Q}_1
            + \delta^{\scriptscriptstyle Q}_2)^2
            +(\delta^{\scriptscriptstyle Q}_2
            + \delta^{\scriptscriptstyle Q}_3)^2
            +(\delta^{\scriptscriptstyle Q}_3
            + \delta^{\scriptscriptstyle Q}_1)^2],
           \nonumber \\
   h_x &=& \frac{\nu E_J}{6}\,
            [2(\delta^{\scriptscriptstyle \Phi}_2
            +\delta^{\scriptscriptstyle \Phi}_3)^2
            -(\delta^{\scriptscriptstyle \Phi}_1
            +\delta^{\scriptscriptstyle \Phi}_2)^2
            -(\delta^{\scriptscriptstyle \Phi}_1
            +\delta^{\scriptscriptstyle \Phi}_3)^2]
            \nonumber \\ &&
           +\frac{t \pi^2}{6}\,
           [2{\delta^{\scriptscriptstyle Q}_1}^2
           -{\delta^{\scriptscriptstyle Q}_2}^2
           -{\delta^{\scriptscriptstyle Q}_3}^2
           -\,2\delta^{\scriptscriptstyle Q}_3
           (\delta^{\scriptscriptstyle Q}_2
           -\delta^{\scriptscriptstyle Q}_1)
           \nonumber \\ &&
           \qquad\qquad\qquad\qquad\qquad\quad
           -\,\,2\delta^{\scriptscriptstyle Q}_2
           (\delta^{\scriptscriptstyle Q}_3
           -\delta^{\scriptscriptstyle Q}_1)],
           \nonumber \\
   h_y &=& \frac{\nu E_J}{2\sqrt{3}}\,
           (2\delta^{\scriptscriptstyle \Phi}_1
           +\delta^{\scriptscriptstyle \Phi}_2
           +\delta^{\scriptscriptstyle \Phi}_3)
           (\delta^{\scriptscriptstyle \Phi}_2
           -\delta^{\scriptscriptstyle \Phi}_3)
           \nonumber \\ &&
           +\frac{t \pi^2}{2\sqrt{3}}\,
           [{\delta^{\scriptscriptstyle Q}_3}^2
           -{\delta^{\scriptscriptstyle Q}_2}^2
           +2\delta^{\scriptscriptstyle Q}_1
           (\delta^{\scriptscriptstyle Q}_3
           -\delta^{\scriptscriptstyle Q}_2)],
           \nonumber \\
   h_z &=& -\frac{\pi\, s t}{\sqrt{3}}\,
           [(\delta^{\scriptscriptstyle\Phi}_1
           +\delta^{\scriptscriptstyle\Phi}_2)
           (\delta^{\scriptscriptstyle Q}_1
           +\delta^{\scriptscriptstyle Q}_2)
           \label{fieldsdd} \\ &&
           +(\delta^{\scriptscriptstyle\Phi}_2
           +\delta^{\scriptscriptstyle\Phi}_3)
           (\delta^{\scriptscriptstyle Q}_2
           +\delta^{\scriptscriptstyle Q}_3)
           +(\delta^{\scriptscriptstyle\Phi}_3
           +\delta^{\scriptscriptstyle\Phi}_1)
           (\delta^{\scriptscriptstyle Q}_3
           +\delta^{\scriptscriptstyle Q}_1)].
           \nonumber
\end{eqnarray}
For convenience, we also cite the energy shift $e_{\rm s}$ of
the singlet state,
\begin{eqnarray}
            e_{\rm s}
            &=& \frac{1}{3}
            [v_{1}+v_{2}+v_{3}
           +2(t^s_{12}+t^s_{23}+t^s_{31})],
            \label{des}  \\
            &=& -\frac{4\nu E_J}{3}\,
            ({\delta^{\scriptscriptstyle \Phi}_1}^2
            +{\delta^{\scriptscriptstyle \Phi}_2}^2
            +{\delta^{\scriptscriptstyle \Phi}_3}^2
            +\delta^{\scriptscriptstyle \Phi}_1
            \delta^{\scriptscriptstyle \Phi}_2
            +\delta^{\scriptscriptstyle \Phi}_1
            \delta^{\scriptscriptstyle \Phi}_3
            +\delta^{\scriptscriptstyle \Phi}_2
            \delta^{\scriptscriptstyle \Phi}_3)
           \nonumber \\ &&
           -\frac{t\pi^2}{3}\,
           [(\delta^{\scriptscriptstyle Q}_1
            +\delta^{\scriptscriptstyle Q}_2)^2
            +(\delta^{\scriptscriptstyle Q}_2
            +\delta^{\scriptscriptstyle Q}_3)^2
            +(\delta^{\scriptscriptstyle Q}_3
            +\delta^{\scriptscriptstyle Q}_1)^2].
            \nonumber
\end{eqnarray}
While the bias variables $\delta^{\scriptscriptstyle \Phi}_i$ and
$\delta^{\scriptscriptstyle Q}_i$ as well as the fields $e_{\rm
d,s}$ and $h_z$ do respect the symmetry of the tetrahedron, the
bias fields $h_x$ and $h_y$ do not. The tetrahedral symmetry may
be made explicit by reexpressing the planar field ${\bf h}_\perp$
in the hexagonal basis ${\bf e}_1 = (1,0)$, ${\bf e}_2 =
(-1/2,-\sqrt{3}/2)$, ${\bf e}_3 = (-1/2,\sqrt{3}/2)$, ${\bf
h}_\perp = \sum_i h_i {\bf e}_i$ with
\begin{equation}
   h_i = \frac{\nu E_J}{3}\,
            (\delta^{\scriptscriptstyle \Phi}_j
            +\delta^{\scriptscriptstyle \Phi}_k)^2\!
           +\frac{t \pi^2}{3}\,
           [{\delta^{\scriptscriptstyle Q}_i}^2\!\!
           +2\delta^{\scriptscriptstyle Q}_i
           (\delta^{\scriptscriptstyle Q}_j
           +\delta^{\scriptscriptstyle Q}_k)^2];
   \label{hi}
\end{equation}
the relation between the tetrahedral symmetries and the rotations
in spin space then becomes obvious. The three symmetric fields
$h_i$ relate to the two cartesian fields via $h_x=h_1-(h_2+h_3)/2$
and $h_y=\sqrt{3}(h_2-h_3)/2$; there is no inverse transformation.

\section{Noise sensitivity and manipulation}
The above results exhibit two remarkable features of our
tetrahedral qubit:

{\it i)} Both, charge and flux noise appear only to quadratic
order, a feature which is easily traced back to the absence
of polarization charges and currents in the qubit's
ground state. This guarantees for a long decoherence
time of the tetrahedral qubit, similar to the
`quantronium' discussed by Vion {\it et al.} \cite{vion_02}.
Numerical analysis confirms the weak susceptibility to charge
noise: a uniform random bias $V \in [-V_0,V_0]$ acting on the
islands produces a small doublet splitting $\delta \approx 0.2\,
\Delta_{\rm d} (2eV_0/E_C)^2$, roughly independent of $E_J/E_C$.
At the same time, the quadratic dependence on bias allows for a
qubit manipulation via $ac$-fields and the dangerous low-frequency
noise can be blocked with appropriate filters.

{\it ii)} The tetrahedral qubit admits a large variety of
manipulation schemes using either magnetic or electric bias.
The `planar fields' $h_{x}$ and $h_y$ can be manipulated
via changes in flux alone, e.g., setting
$\delta^{\scriptscriptstyle \Phi}_2=
\delta^{\scriptscriptstyle \Phi}_3 = \delta^{\scriptscriptstyle
\Phi} \neq \delta^{\scriptscriptstyle \Phi}_1/a_x$ with $a_x =
(-1\pm 2)$, we can direct the field along the $x$-axis, while the
flux state $\delta^{\scriptscriptstyle \Phi}_1 =
\delta^{\scriptscriptstyle \Phi}[(1+a_y^2)/2]^{1/2}$,
$\delta^{\scriptscriptstyle \Phi}_2/a_y =
\delta^{\scriptscriptstyle \Phi}_3 = \delta^{\scriptscriptstyle
\Phi}$, with $a_y= -(2+\sqrt{3})$ produces a field pointing in the
$y$-direction (here, we assume no charge bias,
$\delta^{\scriptscriptstyle Q}_i = 0$; with these choices of
parameters the subleading term $E_J ({\delta^{\scriptscriptstyle
\Phi}_1}^2 +{\delta^{\scriptscriptstyle \Phi}_3}^2
-{\delta^{\scriptscriptstyle \Phi}_2}^2)/2$ in (\ref{dH})
contributing to $h_x$ with $(E_J/6) [{\delta^{\scriptscriptstyle
\Phi}_2}^2+ {\delta^{\scriptscriptstyle \Phi}_3}^2-
2{\delta^{\scriptscriptstyle \Phi}_1}^2]$ and to $h_y$ with
$(E_J/2\sqrt{3}) [{\delta^{\scriptscriptstyle \Phi}_2}^2-
{\delta^{\scriptscriptstyle \Phi}_3}^2]$, vanish as
well). Proper choice of amplitudes and phases allows for the
generation of a rotating planar field. The axial field $h_z$,
however, involves a modification of fluxes {\it and} charges, cf.\
(\ref{fieldsdd}); choosing a uniform flux-
$\delta^{\scriptscriptstyle \Phi}_i =\delta^{\scriptscriptstyle
\Phi}$ and a uniform charge-bias $\delta^{\scriptscriptstyle Q}_i
=\delta^{\scriptscriptstyle Q}$ produces a pure axial field
\begin{equation}
  h_z = -4\pi \sqrt{3} s t \delta^{\scriptscriptstyle \Phi}
  \delta^{\scriptscriptstyle Q}.
  \label{hz}
\end{equation}

The free manipulation of the `magnetic field' ${\bf h}(t)$ allows
for the implementation of Berry-phase type phenomena
\cite{berry_84,wilczek_84,shapere_89}. E.g., the following
provides a simple realization of the NOT-operator in the basis
$|s,a\rangle = [|+\rangle \pm |-\rangle]/{\sqrt{2}}$:
adiabatically rotating the transverse components $h_{x,y} =
h_\perp \exp(i\omega_{h_\perp} t)$, while keeping the $z$-component
$h_z$ fixed, defines the operator $\hat{U}_{\rm Berry} =
\exp{i\sigma_z \Omega/2}$ after one period of rotation. Here,
$\Omega = 2\pi (1-h_z/\sqrt{h_z^2 + h_\perp^2})$ is the solid
angle spanned by the rotating field-cone. Selecting a field vector
$\bf h$ pointing at an angle of $60^\circ$ with respect to the
$z$-axis, the resulting operator $\exp(i\pi\sigma_z/2)$ changes
the relative sign of the components along $|\pm \rangle$, i.e.,
the eigenstates $|s,a\rangle$ of the operator $\sigma_x$ transform
into one another upon each period of field rotation. Note that
operator $\hat{U}_{\rm Berry}$ does not depend upon the rotating
field frequency $\omega_{h_\perp}$ as long as $\omega_{h_\perp}^{-1}$ is much
shorter than the qubit's decoherence time $t_{\rm dec}$.

\section{Measurements}
Finally, we discuss several potential procedures for the
measurement of our qubit's state. We assume an ideal symmetric
device. As the qubit reacts to external bias only in second order,
the measurement of its state involves a two-step process: In a
first step, the qubit is pushed away from the symmetric point of
operation through appropriate driving via external charge
($\delta^{\scriptscriptstyle Q}_i$) and flux
($\delta^{\scriptscriptstyle \Phi}_i$) bias. As a consequence, the
previously quiet qubit state now develops finite internal currents
and polarization charges. In a second step, these signals have to
be measured by appropriate devices and the meter reading will tell
about the qubit's state.

\begin{figure}[ht]
\includegraphics[scale=0.35]{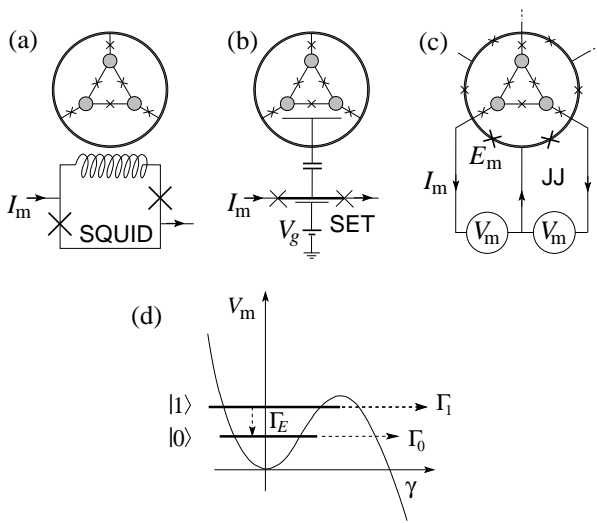}
\caption[]{Measurement setups allowing the identification of the
  qubit state. (a) An external charge bias induces currents in the
  qubit structure with a circularity depending on the qubit state.
  The flux associated with these currents is inductively coupled to
  a SQUID which is driven close to criticality during the
  measurement. Depending on the qubit flux, the SQUID is driven
  overcritical and the associated voltage is measured. (b) An
  external current bias induces polarization charges in the qubit
  structure with a polarity depending on the qubit state. The
  associated charge is capacitively coupled to a SET which is driven
  close to (charge) frustration during the measurement. Depending on
  the qubit's polarization charge the SET is driven into the
  conducting state and the associated current is measured. (c)
  Similar to (a), but with qubit currents directly channelled
  through the measurement junctions with large couplings
  $E_\mathrm{m} \gg E_{J}$; again, the presence of a voltage
  $V_\mathrm{m}$ in the external loop carries the information on the
  qubit state. (d) States of the measurement junctions (with phase
  $\gamma$) at fixed classical driving current $I_\mathrm{m}$ 
  with a slowly decaying state $|0\rangle$ (decay rate $\Gamma_0$) 
  and a fast decaying state $|1\rangle$ (decay rate $\Gamma_1 
  \gg \Gamma_0$) depending on the qubit state. We assume a slow 
  energy relaxation for the qubit, $\Gamma_E \ll \Gamma_1$.}
  \label{fig:tetrahedron_4}
\end{figure}

As a simple illustration we consider the measurement of the
qubit's state in the basis $|\pm\rangle$, i.e., the operator
$\sigma_z$, using a charge bias $\delta^{\scriptscriptstyle
Q}_i \ll 1$. This charge bias generates a current flow within the
qubit and the associated flux can be measured by an external SQUID
loop \cite{robertson_02}, cf.\ Fig. \ref{fig:tetrahedron_4}(a).
We express the matrix (\ref{dH_O}) in the basis
$\{|+\rangle,|-\rangle,|0\rangle\}$,
\begin{equation}
   \delta H_{O} = \left(
         \begin{array}{ccc}
         e_{\rm d}+h_z & h_x - i h_y  & e^+_{0}+ih^+_{0y}\\
         h_x+ih_y      &e_{\rm d}-h_z & e^-_{0}-ih^-_{0y}\\
         e^+_{0}-ih^+_{0y}& e^-_{0}+ih^-_{0y}& e_{\rm s}
         \end{array}
         \right),
   \label{dH_O_pm0}
\end{equation}
with $e_{\rm d,s}$, $h_x,h_y,h_z$ given in (\ref{fields}) and
(\ref{des}) above and
\begin{eqnarray}
   e^\pm_{0} &=& \frac{1}{6}[2v_{1}-v_{2}-v_{3}
             -(2t^s_{23}-t^s_{12}-t^s_{31})]
             \nonumber \\
             &&\pm \frac{1}{2\sqrt{3}}(2t^a_{23}-t^a_{12}-t^a_{31}),
             \label{fields0} \\
   h^\pm_{0y} &=& \frac{1}{2\sqrt{3}}[v_{3}-v_{2}
             -(t^s_{12}-t^s_{31})]
             \pm\frac{1}{2} (t^a_{12}-t^a_{31}).\nonumber
\end{eqnarray}
A finite `magnetic field' ${\bf h} = (0,0,h_z)$ induces the shifts
$\delta E^\pm = \pm h_z $ in the two states $|\pm\rangle$.
Charge-biasing the device induces the current (we assume 
$\delta^{\scriptscriptstyle Q}_i\ll 1$)
\begin{eqnarray}
   I_i^\pm &=& \frac{2e}{\hbar} \frac{\partial \delta E^\pm}
   {\partial \delta^{\scriptscriptstyle \Phi}_i}
   \label{current}\\
   &=& \mp \frac{2 \pi e s t}{\sqrt{3}\hbar}
   (2\delta^{\scriptscriptstyle Q}_i
   +\delta^{\scriptscriptstyle Q}_j
   +\delta^{\scriptscriptstyle Q}_k)\nonumber
   \end{eqnarray}
in the corresponding sub-loop `$0-j-k$' (the above signs apply to
the isolated tetrahedron and have to be reversed for the connected
device; a positive current runs counter clockwise around the
loop). Alternatively, applying a (small) external flux-bias
induces the voltage (see Fig.\
\ref{fig:tetrahedron_4}(b))
\begin{eqnarray}
   V_i^\pm &=& \frac{1}{2e}\frac{\partial \delta E^\pm}
           {\partial\delta^{\scriptscriptstyle Q}_i}
           \label{voltage}\\
           &=& \mp \frac{\pi s t}{2e \sqrt{3}}
           (2\delta^{\scriptscriptstyle \Phi}_i
           +\delta^{\scriptscriptstyle \Phi}_j
           +\delta^{\scriptscriptstyle \Phi}_k)
           \nonumber
\end{eqnarray}
on the island `$i$' which can be measured via a single electron
transistor (SET) device \cite{aassime_01}; here $i,j,k \in
\{1,2,3\}$ and pairwise different. Choosing the bias in the ratio
$\delta_i^{\scriptscriptstyle \Phi, Q} :
\delta_j^{\scriptscriptstyle \Phi, Q} :
\delta_k^{\scriptscriptstyle \Phi, Q} = 3 : -1 : -1$ limits the
current/voltage to the loop/island `$i$'.

Several issues have to be considered in the measurement process,
cf.\ the discussion in Ref.\ \onlinecite{schoen_review}. Relevant
parameters are the dephasing- and mixing rates $\Gamma_\varphi$
and $\Gamma_E$ induced by the measurement apparatus and their
relation to the separation $E_{01}$ between the qubit eigenstates;
in a weak measurement scheme we have $\Gamma_\varphi \ll E_{01}$,
while a projective measurement is characterized by a strong 
coupling with $E_{01} \ll \Gamma_\varphi$ such that the quantum 
evolution of the system is quenched rapidly. While dephasing 
transforms a coherent superposition of states into
a classical mixture through elimination of the off-diagonal
elements in the qubit's density matrix, the mixing induces
transitions between the qubit's states and thus spoils the
measurement. Usually, a good measurement setup makes use of
decoherence in transferring classical information to the
measurement device but avoids mixing, hence $\Gamma_\varphi \gg
\Gamma_E$. 

The simplest situation is realized when the measured 
observable commutes with the qubit Hamiltonian --- in this case 
the measurement preserves the qubit's eigenstates. On the
other hand, if the measured observable does not commute with 
the qubit Hamiltonian, the measurement has to be completed before 
mixing sets in; hence $\Gamma_E<\Gamma_\mathrm{meas}<\Gamma_\varphi
\ll E_{01}/\hbar$ for a weak measurement, while the
sequence $\Gamma_E < \Gamma_\mathrm{meas} < E_{01}/\hbar \ll
\Gamma_\varphi$ applies to the projective measurement (here,
$\Gamma_\mathrm{meas} = 1/t_\mathrm{meas}$ denotes the inverse
measuring time). 
\begin{figure}[ht]
\includegraphics[scale=0.30]{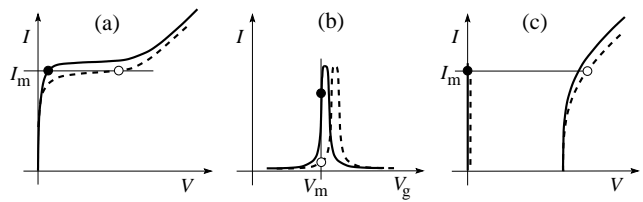}
\caption[]{Current-voltage characteristics of meter devices.
  The solid/dashed lines refer to the different quantum states
  of the qubit shifting the characteristic of the meter device.
  (a)
  In an overdamped SQUID the current-voltage characteristic is
  single valued, with a voltage due to a finite rate of individual
  phase-slips. $I_\mathrm{m}$ is the imposed measuring curent.
  (b)
  The current in the SET results from a continuous flow of
  individual electrons traversing the island. $V_\mathrm{m}$ is
  the imposed gate voltage during measurement.
  (c)
  An underdamped SQUID/Josephson junction exploits an
  instability where a single phase-slip triggers a transition to
  the dissipative branch.}
  \label{fig:tetrahedron_5}
\end{figure}

The measurement schemes described in Ref.\ 
\onlinecite{schoen_review} involve dissipative meter devices, 
e.g., an overdamped SQUID with a well defined (single-valued)
current-voltage characteristic as depicted in Fig.\
\ref{fig:tetrahedron_5}(a) or a SET with the characteristic shown
in Fig.\ \ref{fig:tetrahedron_5}(b). In both cases the measurement
involves numerous dissipative events, either many phase-slips
producing the voltage across the SQUID device or many electrons
traversing the island of the SET. The fluctuations due to 
the phase slips/electrons act back on the qubit, enforcing
its loss of phase coherence. Such measurement schemes can be 
implemented in terms of weak or strong (projective) measurements.

\begin{figure}[ht]
\includegraphics[scale=0.30]{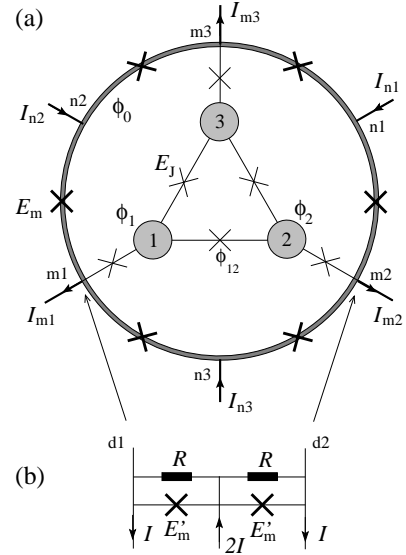}
\caption[]{(a) Symmetric setup measuring the quantum state of the
  qubit. Currents below twice the critical current $I_{\mathrm{m},c}
  = 2e E_\mathrm{m}/\hbar$ of the six equal measurement junctions
  are fed into the qubit loop through the points `n1', `n2', and `n3',
  and removed through the leads at `m1', `m2', and `m3'. The qubit current
  $I_{\mathrm{q}}$ flowing in the segments `n3-m1' and `n3-m2' (and 
  cyclically permutated) add to these measurement currents and drive 
  the junction towards or away from criticality. The voltage pattern 
  measured on the six junctions tells about the qubit's state.
  Care has to be taken not to spoil the symmetry in this
  measurement setup through a ring current in the outer loop; such
  a current can be avoided by appropriate compensation of the flux
  bias through the individual loops `m$j$-m$k$-$k$-$j$-m$j$' with 
  an equal and opposite flux through the central loop `1-2-3-1'.
  This can be realized, e.g., via an independent control on the individual 
  loops and on the overall flux penetrating the entire structure.
  (b) circuitary introducing damping on the six measuring junctions.
  Driving the parallel junctions (with couplings $E_J^\prime$)
  overcritical via a large current $2I$ switches the parallel
  channel into the resistive state. The contacts `d1' and `d2' of
  such a dissipative element are then connected to the points `m1', 
  `m2', and `m3' of the qubit structure in order to make the
  associated currents classical.}
  \label{fig:tetrahedron_6}
\end{figure}

This type of measurement has to be contrasted with a  
meter characterized by an instability, such as an underdamped
SQUID or Josephson junction with a characteristic as shown in
Fig.\ \ref{fig:tetrahedron_5}(c). Such a meter does not couple
dissipatively to the qubit but switches to a dissipative state
only after the measurement, e.g., after the occurrence of one
phase slip, hence $\Gamma_\varphi \ll E_{01}$; the measurement
is weak and generates a small imaginary part to the energy of 
the qubit eigenstate which then may be determined in a decay 
process. The qubit states are identified through their 
(exponentially) different decay rates. Such a measurement 
scheme has been used recently by Vion {\it et al.} \cite{vion_02}. 
In their setup, an additional measurement junction is introduced 
into the qubit loop. The current generated by the qubit is 
superimposed on an external measurement current and drives 
the measurement junction  towards criticality, see Fig.\ 
\ref{fig:tetrahedron_4}(c). Here, we make use of the 
symmetric setup shown in Fig.\ \ref{fig:tetrahedron_6}(a)
involving six classical measurement junctions with equal couplings
$E_\mathrm{m} \gg E_J$. The measuring currents $I_{\mathrm{m}i}$
are fed into the device through the points `n1', `n2', and `n3'
and removed at the points `m1', `m2', and `m3'. The measuring
current $I_\mathrm{m}$ flowing in the segments `n3-m1' and `n3-m2'
(and in the other obtained through cyclic permutation) is chosen
close to the critical current $I_{\mathrm{m},c} = 2e
E_\mathrm{m}/\hbar$ of the measurement junctions, requiring to
feed a current $I_{\mathrm{ext}} \approx 2 I_{\mathrm{m},c}$ into the
tetrahedron via the input lines `n1', `n2', `n3' and retraining them
via the output lines `m1', `m2', `m3'. A convenient measuring
setup is the symmetric one with equal currents crossing the
junctions `n3-m1' and `n3-m2' (and other pairs obtained through
cyclic permutation). Driving the qubit with external bias fields
$\delta^{\scriptscriptstyle \Phi}_l$ and
$\delta^{\scriptscriptstyle Q}_l$ induces additional currents
$I_{\mathrm{q},i}$ in the loops `m$j$-m$k$-$k$-$j$-m$j$' which are
characteristic for the quantum state of the qubit. Depending on
the relative flow direction of the qubit currents and the
measuring current, the six measuring junctions are either driven
towards ($I_\mathrm{m} + I_{\mathrm{q},i} \sim I_{\mathrm{m},c}$)
or away ($I_\mathrm{m} + I_{\mathrm{q},i} < I_{\mathrm{m},c}$)
from criticality. The voltage pattern appearing on the six
junctions then allows for the identification of the qubit state.
Within this scheme, care has to be taken not to spoil the symmetry
of the measurement setup by a circular current flowing in the
outer ring, hence the system should be flux biased such that the
total flux $\Phi$ through the ring remains zero. This can be
easily achieved by compensating the fluxes $\Phi_i$ in the loops
`m$j$-m$k$-$k$-$j$-m$j$' through an equal and opposite flux
$\Phi_\triangle = -\sum_{i=1}^3 \Phi_i$ through the central
triangular loop. 

Below, we study two schemes, a weak measurement in the Hamiltonian
basis with $\Gamma_\varphi < E_{01}$ and a projective measurement
onto the current basis with $E_{01} < \Gamma_\varphi$. In both cases
we discuss the measurement of two operators, the measurement of
$\sigma_z$ and of $\sigma_x$.

\subsection{Measurements in the qubit's eigenbasis}
A proper measurement in the energy eigenbasis imposes a number of
constraints on the measuring device. First of all, the phase
decoherence rate has to be small, $\Gamma_\varphi \ll E_{01}$. The
decay rate $\Gamma_1$ via tunneling of the high energy state (low
barrier state, cf.\ Fig.\ \ref{fig:tetrahedron_4}(d)) should be
large, $\Gamma_1 \gg \Gamma_E$ such that the system tunnels before
decaying to the (metastable) ground state. Finally, the
measurement time $t_\mathrm{meas}$ should be larger than the
inverse tunneling rate, $t_\mathrm{meas} > \Gamma_1^{-1}$. Note,
that phase decoherence before tunneling, i.e., $\Gamma_\varphi >
\Gamma_1$, is not a necessary requirement in this type of
measurement; the decay may as well proceed out of the coherent
state, i.e., a superposition of the qubit states. In this case,
the first phase slip triggers the projection, while the subsequent
phase-slips produce the large voltage signal.

In order to measure the operator $\sigma_{\bf n}$, i.e., the spin
projection onto the axis ${\bf n}$, we first apply a `magnetic
field' ${\bf h} = h {\bf n}$ directed along ${\bf n}$. This is
achieved via proper charge- and flux-biasing as described by the
equations (\ref{fieldsdd}). The doublet space $\{|+\rangle,
|-\rangle\}$ then is split with new qubit eigenstates
$|+\rangle_{\bf n}$ and $|-\rangle_{\bf n}$ separated by the
`Zeeman' energy $\Delta E = 2h$. The currents $I^{\rm
\scriptscriptstyle (n{\it j}-m{\it k})}$ flowing in the segments
`m$j$-n$k$' across the measuring junctions are equal to the loop
currents $I_{\mathrm{q},i}$ in `m$j$-m$k$-$k$-$j$-m$j$' and thus
are determined by the derivatives
\begin{equation}
  I^{\rm \scriptscriptstyle (n{\it j}-m{\it k})} = I_{\mathrm{q},i}
  = - \frac{2e}{\hbar} \frac{\partial h}{\partial
  \delta^{\scriptscriptstyle \Phi}_i}.
  \label{I_minj}
\end{equation}
In order to avoid the flow of a circular current in the outer
ring, the driving flux bias $\delta^{\scriptscriptstyle \Phi}_i$
has to be properly compensated as described above.

We proceed with the evaluation of the currents (\ref{I_minj}) 
associated with the measurement of $\sigma_z$ and $\sigma_y$
and generated by the application of magnetic fields $h_z$ and
$h_x$ along the $z$- and $x$-axis, respectively. Starting with the
measurement of $\sigma_z$, the projection along the $z$-axis, we
choose a uniform charge and flux bias ($\rightarrow {\bf h}
= (0,0,h_z)$, cf.\ (\ref{fieldsdd})) and obtain the qubit loop 
currents
\begin{equation}
  I_{\mathrm{q},i}^z = \pm \frac{8 \pi e s t}{\sqrt{3}\hbar}
  \delta^{\scriptscriptstyle Q},
  \label{Iqz1}
\end{equation}
$i=1,2,3$, in the qubit state $|\pm\rangle_z$. Hence all loops 
are equally driven and voltage signals on the triple `n1-m3', 
`n2-m1', `n3-m2' identify the qubit state $|+\rangle$, while 
finite voltages on the junctions `n1-m2', `n2-m3', `n3-m1' 
are associated with the $|-\rangle$ state, provided that 
$\delta^{\scriptscriptstyle Q}>0$.

Second, the projection along the $x$-axis can be measured by
applying a `field' along $h_x$ using a flux bias
$\delta_2^{\scriptscriptstyle \Phi} = \delta_3^{\scriptscriptstyle
\Phi} = - \delta_1^{\scriptscriptstyle \Phi} \equiv
\delta^{\scriptscriptstyle \Phi}$. This bias generates a field
$h_x = 4 \nu E_J (\delta^{\scriptscriptstyle \Phi})^2/3$ and
imprints qubit currents of magnitude 
\begin{equation}
  I_{\mathrm{q},i}^x = \mp \frac{8e\nu E_J}{3\hbar}
  \delta^{\scriptscriptstyle \Phi}
  \label{Iqz2}
\end{equation}
for $i=2,3$ in the qubit state $|\pm\rangle_x$, while 
$I_{\mathrm{q},1}^x = 0$. Hence the pairs `n2-m1', `n3-m2', 
and `n2-m3', `n3-m1' are equal and oppositely driven, while 
the junctions `n1-m2' and `n1-m3' do not experience
any additional drive due to the qubit structure.

\subsection{Projective measurement in the current basis}
A projective measurement in the current basis requires to switch
on a strong decoherence $\Gamma_\varphi \gg E_{01}$ during the
measurement. This decoherence then projects the state of the qubit
at onset of dissipation onto the current basis and keeps it there
via the Zeno (watchdog) effect, cf.\ Ref.\
\onlinecite{schoen_review}. A suitable circuit allowing to turn on
decoherence is shown in Fig.\ \ref{fig:tetrahedron_6}(b). The
admittance between the points `d1' and `d2' is given by the
expression $Z^{-1}(\omega)= 1/2R - E_\mathrm{m}^\prime
/2i\hbar\omega R_Q$, with $R_Q = \hbar/4e^2$ the quantum
resistance. Choosing parameters $R \lesssim R_Q$ (this guarantees
a sufficient decoherence in the `on' state, see below),
$E_\mathrm{m}^\prime \sim (10 - 100) E_J$, and an operating
frequency $\omega \lesssim 0.1 E_J$, we find that $Z 
\approx -2iR_Q (\hbar\omega/E_\mathrm{m}^\prime)$.
Hence, at zero applied current $I$ the conductance between the
points `d1' and `d2' is dominated by the Josephson junctions and
is mostly imaginary at low frequencies, allowing us to ignore
dissipation (`off' state). However, when the system is biased 
by a current $I$ larger than critical, the Josephson current 
disappears and the resistance $R$ provides a significant source 
of dissipation as quantified through the dimensionless parameter
$\alpha = R_Q/R$. In addition, while the Josephson junction itself 
involves a large quasi-particle resistance at low temperatures, 
when driven with a large current the junction switches to the 
resistive state with the resistance $R_\mathrm{m}^\prime$;
the latter is related to $E_\mathrm{m}^\prime$ via the
Ambegaokar-Baratoff relation $R_\mathrm{m}^\prime/R_Q=\Delta/
E_\mathrm{m}^\prime$. The projective measurement in the 
current basis then starts with switching on a strong 
dissipation (such that $\Gamma_\varphi \gg E_{01}$) 
which is diagonal in the eigenbasis of the qubit's current
operator $\hat{I}_{\mathrm{q}}$.

Before continuing with the discussion of an appropriate
projective measurement in the current basis, we discuss
one more issue related to the quality of such a measurement.
The above measurement schemes (via coupling to a SQUID or a SET)
are diagonal in the qubit's subspace spanned by $|+\rangle$ and
$|-\rangle$ and can be implemented in terms of weak and strong
(projective) measurements, cf.\ Ref.\ \onlinecite{schoen_review}.
However, extending the analysis to the low-energy subspace
$\{|+\rangle,|-\rangle, |0\rangle\}$, we find for the current
operator $\hat{I}_1= (2e/\hbar) \partial \delta H_{O}/\partial
{\delta^{ \scriptscriptstyle \Phi}_1}|_{\delta^{\scriptscriptstyle
\Phi}_1=0}$ the expression (cf.\
(\ref{dH_O_pm0}))
\begin{eqnarray}
    \frac{2\pi est}{\hbar\sqrt{3}}\! \left(
         \begin{array}{ccc}
         \!-(\delta^{\scriptscriptstyle Q}_{12}
         +\delta^{\scriptscriptstyle Q}_{31}) & 0  &
         \delta^{\scriptscriptstyle Q}_{12}\zeta
         +\delta^{\scriptscriptstyle Q}_{31}\zeta^*\!\! \\
         \!0 &\delta^{\scriptscriptstyle Q}_{12}
         +\delta^{\scriptscriptstyle Q}_{31} &
         -\delta^{\scriptscriptstyle Q}_{12}\zeta^*
         -\delta^{\scriptscriptstyle Q}_{31}\zeta\!\! \\
          \!\delta^{\scriptscriptstyle Q}_{12}\zeta^*
          +\delta^{\scriptscriptstyle Q}_{31}\zeta &
          -\delta^{\scriptscriptstyle Q}_{12}\zeta
          -\delta^{\scriptscriptstyle Q}_{31}\zeta^*& 0
         \!\! \end{array}
         \right)\!,\nonumber
\end{eqnarray}
where $\delta^{\scriptscriptstyle Q}_{ij} =
\delta^{\scriptscriptstyle Q}_i+ \delta^{\scriptscriptstyle Q}_j$;
choosing a charge-bias $\delta^{\scriptscriptstyle Q}_1/3 =
\delta^{\scriptscriptstyle Q} = -\delta^{\scriptscriptstyle Q}_2 =
-\delta^{\scriptscriptstyle Q}_3$ this reduces to
\begin{eqnarray}
    \hat{I}_1=
    -\frac{4\pi est}{\hbar\sqrt{3}}\delta^{\scriptscriptstyle Q}
         \left(
         \begin{array}{ccc}
         2 &  0 &  1\\
         0 & -2 & -1\\
         1 & -1 &  0
         \end{array}
         \right).\label{smeas}
\end{eqnarray}
Similar expressions apply to the other current operators
$\hat{I}_2$ and $\hat{I}_3$ and, replacing $2\pi est/\hbar\sqrt{3}
\rightarrow \pi st/2e\sqrt{3}$ in the prefactor and substituting
$\delta^{\scriptscriptstyle Q}_{ij}\rightarrow
\delta^{\scriptscriptstyle \Phi}_{ij} = \delta^{\scriptscriptstyle
\Phi}_i+ \delta^{\scriptscriptstyle \Phi}_j$, to the voltage
operators $\hat{V}_i = (1/2e) \partial\delta H_{O}/\partial
{\delta^{\scriptscriptstyle Q}_i}$.

Hence a projective measurement onto the current basis is in fact 
non-ideal as the measured observable, the current, does not 
commute with the Hamiltonian when going beyond the qubit sector. 
In a `high quality' measurement of our qubit state one would 
request that the off-diagonal matrix elements (between the 
doublet and singlet states) of the operator are much smaller 
than the matrix elements in the doublet subspace. Otherwise, 
the measurement has to be repeated many times in order
to arrive at a proper readout of the qubit state. 
We then need to identify special measurement configurations 
where the off-diagonal matrix elements remain small or even 
vanish altogether. 

We will now give a specific example how this goal can be 
achieved for a projective measurement of the $\sigma_z$ 
operator. We will show that a charge-bias with
\begin{equation}
   \delta^{\scriptscriptstyle Q}_2
   = - \delta^{\scriptscriptstyle Q}_3 =
   \delta^{\scriptscriptstyle Q}, \qquad
   \delta^{\scriptscriptstyle Q}_1 = 0
   \label{Q_bias}
\end{equation}
results in a high-quality measurement of $\sigma_z$ and $\sigma_y$
if we choose a large bias $\delta^{\scriptscriptstyle Q} = 1/2$.
This bias induces a current flow of equal magnitude
but different chirality in the loops (cf.\ Fig.\ 
\ref{fig:tetrahedron_6}(a))
`m1-1-3-m3-m1' (current $I_2$) and `m2-2-1-m1-m2'
(current $I_3$; loop currents circulating counter clockwise are
positive); no current is induced in the loop `m3-3-2-m2-m3'. The
total current through the link `m1-1' is given by $I^{\rm
\scriptscriptstyle (m1-1)}= I_2 - I_3$ and will be used to drive
the measurement current $I_{\rm m}$ fed symmetrically into the 
tetrahedron at the points `n$i$' overcritial. We first describe
the setup projecting the link current $I^{\rm
\scriptscriptstyle (m1-1)}$ and subsequently derive an explicit 
expression for the current operator. In a third step, we then
describe the actual measurement process in detail.

\begin{figure}[ht]
\includegraphics[scale=0.30]{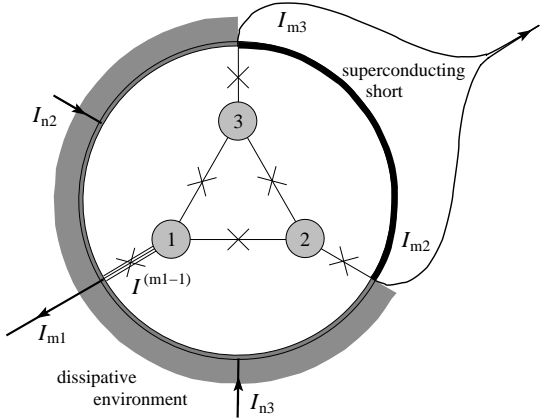}
\caption[]{Dissipative measurement setup projecting
  the qubit state onto the fixed link current $I^{\rm
  \scriptscriptstyle (m1-1)}$.}
  \label{fig:tetrahedron_7}
\end{figure}

We will show below that the link current $I^{\rm
\scriptscriptstyle (m1-1)}$ will not mix to the third state
$|0\rangle$. However, this property is not shared by the 
individual loop currents $I_2$ and $I_3$. It is then important
to devise a setup projecting the state onto the link current
$I^{\rm \scriptscriptstyle (m1-1)}$ rather than the loop
currents $I_2$ or $I_3$. This is achieved by the symmetric 
coupling of three `dissipators' in between the points `m1-m2',
`m2-m3', and `m3-m1'. Turning on one of these dissipators
renders the current in the corresponding link classical.
In our setup we wish both loop current, $I_2$ and $I_3$,
to be classical and hence switch on the dissipators 
`m1-m2' and `m1-m3'. In this situation it is the link
current $I^{\rm \scriptscriptstyle (m1-1)}$ which is 
coupled to the dissipative reservoir (the segment `m2-m3'
remains quantum and can be contracted to a point, cf.\
Fig.\ \ref{fig:tetrahedron_7}) and hence the quantum
state is projected to a state with fixed current 
$I^{\rm \scriptscriptstyle (m1-1)}$. In an ideally
symmetric setup this current will be symmetrically
split into the two arms `m1-m2' and `m1-m3' and,
depending on the qubit state, will drive a specific set
of junctions overcritical. 

Let us then turn to the calculation of the link current
$I^{\rm \scriptscriptstyle (m1-1)}$.
In a first step, we have to generalize the tunneling matrix
elements (\ref{tsa}) to allow for a large charge bias,
\begin{eqnarray}
   t+t^s_{ij} &=&
   t\, \cos(\pi \delta^{\scriptscriptstyle Q}_{ij}),
   \nonumber\\
   t^a_{ij} &=& s\, t\,
   \delta^{\scriptscriptstyle \Phi}_{ij}
   \sin(\pi\delta^{\scriptscriptstyle Q}_{ij}).
   \label{tsa_gen}
\end{eqnarray}
Assuming the above form of charge bias, the Hamiltonian in the
basis of semi-classical states $\{|O_1\rangle,|O_2\rangle,
|O_3\rangle\}$ takes the form $H_O=t h_{Or} +ist h_{Oi}$
with
\begin{eqnarray}
    h_{Or}=
         \left(
         \begin{array}{ccc}
         0 &  \cos(\pi \delta^{\scriptscriptstyle Q})
           &  \cos(\pi \delta^{\scriptscriptstyle Q})\\
         \cos(\pi \delta^{\scriptscriptstyle Q}) &  0 &  1\\
         \cos(\pi \delta^{\scriptscriptstyle Q}) &  1 &  0
         \end{array}
         \right)\label{H_O_lqr}
\end{eqnarray}
and
\begin{eqnarray}
    h_{Oi}\!=\!
         \left(
         \begin{array}{ccc}
         \!\!0 &  \delta_{12}^{\scriptscriptstyle \Phi}
              \sin(\pi \delta^{\scriptscriptstyle Q})
           &  \delta_{31}^{\scriptscriptstyle \Phi}
              \sin(\pi \delta^{\scriptscriptstyle Q})\! \\
         \!\! -\delta_{12}^{\scriptscriptstyle \Phi}
          \sin(\pi \delta^{\scriptscriptstyle Q}) &  0 &  0\! \\
         \!\!-\delta_{31}^{\scriptscriptstyle \Phi}
          \sin(\pi \delta^{\scriptscriptstyle Q}) &  0 &  0 \!
         \end{array}
         \right)\!,\label{H_O_lqi}
\end{eqnarray}
where $\delta_{ij}^{\scriptscriptstyle \Phi} =
\delta_{i}^{\scriptscriptstyle \Phi} + 
\delta_{j}^{\scriptscriptstyle \Phi}$. The link current 
operator then takes the form (after setting 
$\delta^{\scriptscriptstyle \Phi}_{ij}=0$; note the sign change
for the connected tetrahedron, $\hat{I}_i=-(2e/\hbar) \partial 
\delta H_{O}/\partial{\delta^{\scriptscriptstyle \Phi}_i}$)
\begin{eqnarray}
    \hat{I}^{\rm\scriptscriptstyle (m1-1)}=
    \frac{2 est \, i}{\hbar}\sin(\pi \delta^{\scriptscriptstyle Q})
         \left(
         \begin{array}{ccc}
         0 &  -1 &  1\\
         1 &  0  &  0\\
        -1 &  0  &  0
         \end{array}
         \right).\label{Im11_0}
\end{eqnarray}
In measuring the qubit's state, we first drive the tetrahedron
adiabatically towards the measuring point
$\delta^{\scriptscriptstyle Q} = 1/2$. At small values of $\delta
= (\pi \delta^{\scriptscriptstyle Q})^2/2$ the doublet splits and
we find the new eigenvalues $E_{\delta}$ and associated
eigenvectors $|e_\delta\rangle$ (up to normalization),
\begin{eqnarray}
         \begin{array}{c c}
          E_{\delta}/t & |e_\delta\rangle\\ \noalign{\vskip 6 pt}
         -1
               & |a_\delta\rangle = i
               \frac{{|O_2\rangle-|O_3\rangle}}{\sqrt{2}},\\
         \noalign{\vskip 4 pt}
         -1+\frac{4\delta}{3}~~~
               & |s_\delta\rangle \approx
               \frac{{2(1+\delta/3)|O_1\rangle-|O_2\rangle-|O_3\rangle}}
               {\sqrt{6+8\delta/3}},\\
         \noalign{\vskip 4 pt}
         2-\frac{4\delta}{3}~~~
               & |0_\delta\rangle \approx
               \frac{{(1-\delta/3)|O_1\rangle+|O_2\rangle+|O_3\rangle}}
               {\sqrt{3-2\delta/3}}.
         \end{array}
         \label{ed}
\end{eqnarray}
Note that at $\delta=0$ the eigenvectors $|a_0\rangle$ and $|s_0
\rangle$ defining the low-energy qubit subspace correspond to the
(anti-) symmetric combinations $|a_0\rangle =[|+\rangle
-|-\rangle]/\sqrt{2}$ and $|s_0\rangle=[|+\rangle+|-\rangle]
/\sqrt{2}$. Hence starting with a qubit state $|\Psi\rangle =
\psi_+ |+\rangle + \psi_-|-\rangle$ in the $\sigma_z$ basis
$|\pm\rangle$, the amplitudes which evolve adiabatically towards
the measurement point are the combinations $\psi_{s_0}=(\psi_+ +
\psi_-)/\sqrt{2}$ and $\psi_{a_0} = (\psi_+ - \psi_-)/\sqrt{2}$
or, in matrix form,
\begin{eqnarray}
         \left(
         \begin{array}{c}
         \psi_{a_0} \\
         \psi_{s_0} \\
         \psi_{0_0}
         \end{array}
         \right) =
         \hat{T} \left(
         \begin{array}{c}
         \psi_{+} \\
         \psi_{-} \\
         \psi_{0}
         \end{array}
         \right) = \frac{1}{\sqrt{2}}
         \left(
         \begin{array}{c c c}
         1 & -1 & 0\\
         1 &  1 & 0\\
         0 &  0 & 1\\
         \end{array}
         \right)
         \left(
         \begin{array}{c}
         \psi_{+} \\
         \psi_{-} \\
         \psi_{0} \\
         \end{array}
         \right).
         \nonumber
\end{eqnarray}
Driving the bias up to $\delta^{\scriptscriptstyle Q} = 1/2$ the
spectrum and eigenvectors deform into
\begin{eqnarray}
         \begin{array}{c c}
          E_{1/2}/t & |e_{1/2}\rangle\\ \noalign{\vskip 6 pt}
         -1 &  |a_{1/2}\rangle = i[|O_2\rangle-|O_3\rangle]/{\sqrt{2}},\\
         \noalign{\vskip 4 pt}
         0  & |s_{1/2}\rangle=|O_1\rangle,  \\
         \noalign{\vskip 4 pt}
         1 & |0_{1/2}\rangle= [|O_2\rangle+|O_3\rangle]/{\sqrt{2}}.
         \end{array}
         \label{e12}
\end{eqnarray}
In this new basis $\{|a_{1/2}\rangle,|s_{1/2}\rangle, |0_{1/2}
\rangle\}$ the current operator (\ref{Im11_0}) takes the form
\begin{eqnarray}
    \hat{I}^{\rm\scriptscriptstyle (m1-1)}=
    \frac{2\sqrt{2} est}{\hbar}
         \left(
         \begin{array}{ccc}
         0  &  1  &  0\\
         1 &  0  &  0\\
         0  &  0  &  0
         \end{array}
         \right);\label{Im11}
\end{eqnarray}
hence after the adiabatic evolution, the current measures the
$\sigma_x$ operator in the basis (\ref{e12}) and has no matrix
elements with the third state $|0_{1/2}\rangle$. At the same time,
the qubit wave function takes the form $|\Psi\rangle = e^{-i\theta/2}
\psi_{a_0} |a_{1/2}\rangle + e^{i\theta/2}\psi_{s_0}|s_{1/2}\rangle$,
where $\theta = \int dt (E_a-E_s)/\hbar$ denotes the additional
phase picked up in the adiabatic evolution of the qubit amplitudes
$\psi_{s_\delta}$ and $\psi_{a_\delta}$ until the measurement
is performed. In order to reexpress the current (\ref{Im11})
through the original qubit amplitudes $\Psi_\pm$ we define
the matrix
\begin{eqnarray}
         \hat{T}_\theta =
         \frac{1}{\sqrt{2}}
         \left(
         \begin{array}{c c c}
         e^{-i\theta/2} & -e^{-i\theta/2} & 0\\
         e^{i\theta/2} &  e^{i\theta/2} & 0\\
         0 &  0 & 1\\
         \end{array}
         \right)
         \nonumber
\end{eqnarray}
and find the result
\begin{eqnarray}
    \hat{T}_\theta^\dagger
    \hat{I}^{\rm\scriptscriptstyle (m1-1)} \hat{T}_\theta
    =\frac{2\sqrt{2} est}{\hbar}
         \left(
         \begin{array}{ccc}
         \cos\theta  &  i\sin\theta  &  0\\
         -i\sin\theta  &  -\cos\theta  &  0\\
         0  &  0  &  0
         \end{array}
         \right)
         \label{Im11z}
\end{eqnarray}
describing a high quality measurement of the qubit operators $\pm
\sigma_z$ at $\theta=n\pi$ and $\pm\sigma_y$ at $\theta=
(n+1/2)\pi$ (or any component residing in the $y$-$z$ plane for
angles $\theta$ in between). Applying the charge bias
(\ref{Q_bias}) to cyclically permuted islands $1 \rightarrow 2
\rightarrow 3 \rightarrow 1$ produces measurements of
spin-projections in planes rotated by the corresponding angles
$\pm 2\pi/3$.

In the following, we restrict the discussion to the measurement of
$\sigma_z$, i.e., we assume that $\theta = n\pi$ with $n$ even 
in the specific discussion below. Following the
result (\ref{Im11z}), the link current $I^{\rm \scriptscriptstyle
(m1-1)}$ has opposite signs for the two qubit-states $|+\rangle$
and $|-\rangle$ and its measurement allows for the determination
of the qubit's final wave function. In the actual measurement, the
system is driven symmetrically with equal external currents
$I_{{\rm n}i}= I_{\rm m}$ entering the system at the points `n1',
`n2', `n3' (cf.\ Fig.\ 6(a)) and leaving symmetrically through the
points `m1', `m2', and `m3'. The external current $I_{\rm m}$ is
chosen close to, but below twice the critical current of the large
junctions in the ring, $I_{\rm m1} < 2 I_{{\rm m}c} = 4e E_{\rm
m}/\hbar$. Accounting for the additional currents induced in the
tetrahedron under the charge bias $\delta^{\scriptscriptstyle Q} =
1/2$, the currents $I^{\rm\scriptscriptstyle (n2-m1)}$ and 
$I^{\rm\scriptscriptstyle (n3-m1)}$ through the large 
measurement junctions are equal to $(I_{\rm m} \pm 2\sqrt{2} 
est/\hbar)/2$ for the $|\pm\rangle$ states of the doublet, 
while the currents $I^{\rm\scriptscriptstyle (n2-m3)}$ and 
$I^{\rm\scriptscriptstyle (n3-m2)}$ assume the
values $(I_{\rm m} \mp 2\sqrt{2} est/\hbar)/2$ for the same
eigenstates. Note that  in the singlet state the bare measurement
current $I_{\rm m}/2$ flows through the junctions as no current is
induced in the singlet state $|0\rangle$ under the above charge
bias. As demonstrated in Ref.\ \onlinecite{vion_02}, the
measurement current $I_{\rm m}$ and the induced current $2\sqrt{2}
est/\hbar$ can be chosen such that the switching probabilities
${\cal P}_\pm$ into the transient voltage-state of the large
measurement junctions are strongly different for the $|\pm
\rangle$ states. Then the location of voltage pulses on the
junctions `n2-m1' and  `n3-m1' or on the complementary
junctions `n2-m3' and `n3-m2' tells us whether the qubit was
in the $|+\rangle$ or $|-\rangle$ state just before the
measurement. Furthermore, the absence of any voltage pulse is the
signature of the singlet state. 

Finally, we discuss the projective measurement of the operator
$\sigma_x$. Following (\ref{fieldsdd}), a flux bias produces
finite `magnetic fields' $h_{x,y}$ which are bilinear in
$\delta^{\scriptscriptstyle \Phi}_j$; one then may expect that an
appropriate flux bias will produce loop currents which are
diagonal in the basis $|a,s\rangle$, where
\begin{equation}
    |a\rangle = (|+\rangle - |-\rangle)/\sqrt{2}, \quad
    |s\rangle = (|+\rangle + |-\rangle)/\sqrt{2}.
    \label{sa}
\end{equation}
Here, we concentrate on the main contribution (proportional to
$\nu$) originating from the modification $v_j$ in the energies of
the semi-classical minima $|O_j\rangle$. The current operators
$\hat{I}_i = -(2e/\hbar)\partial \delta H_O/\partial
\delta^{\scriptscriptstyle \Phi}_i$ evaluated in the basis
$\{|+\rangle,|-\rangle,|0\rangle \}$ take the form
\begin{equation}
    \hat{I}_i = \frac{2e\,\nu E_J}{3\hbar}
         \left(
         \begin{array}{ccc}
         \epsilon_i & \lambda_i & \lambda_i^\ast\\
         \lambda_i^\ast &  \epsilon_i &  \lambda_i\\
         \lambda_i &  \lambda_i^\ast &  \epsilon_i
         \end{array}
         \right)\label{Ii_Phi}
\end{equation}
with $\epsilon_i = 4(2\delta^{\scriptscriptstyle \Phi}_i +
\delta^{\scriptscriptstyle \Phi}_j + \delta^{\scriptscriptstyle
\Phi}_k)$ and
\begin{eqnarray}
   \lambda_1 &=& (2\delta^{\scriptscriptstyle \Phi}_1
                     +\delta^{\scriptscriptstyle \Phi}_2
                     +\delta^{\scriptscriptstyle \Phi}_3)
                 +i\sqrt{3}(\delta^{\scriptscriptstyle \Phi}_2
                     -\delta^{\scriptscriptstyle \Phi}_3),\nonumber\\
   \lambda_2 &=& -(2\delta^{\scriptscriptstyle \Phi}_3
                     +\delta^{\scriptscriptstyle \Phi}_2
                     -\delta^{\scriptscriptstyle \Phi}_1)
                     +i\sqrt{3}(\delta^{\scriptscriptstyle \Phi}_1
                     +\delta^{\scriptscriptstyle \Phi}_2),\nonumber\\
   \lambda_3 &=& -(2\delta^{\scriptscriptstyle \Phi}_2
                     +\delta^{\scriptscriptstyle \Phi}_3
                     -\delta^{\scriptscriptstyle \Phi}_1)
                     -i\sqrt{3}(\delta^{\scriptscriptstyle \Phi}_1
                     +\delta^{\scriptscriptstyle \Phi}_3).
   \label{lambda}
\end{eqnarray}
Applying a specific bias $\delta^{\scriptscriptstyle
\Phi}_2=\delta^{\scriptscriptstyle \Phi}_3 =
-\delta^{\scriptscriptstyle \Phi}_1 = \delta^{\scriptscriptstyle
\Phi}$ produces a field along $h_x$ (cf.\ (\ref{fieldsdd})) and
induces the current $I^{\rm\scriptscriptstyle (m3-3)} \equiv I_1 -
I_2 = -I^{\rm\scriptscriptstyle (m2-2)}$ in the loop
`m1-m3-3-1-2-m2-m1',
\begin{eqnarray}
    \hat{I}^{\rm\scriptscriptstyle (m3-3)}=
    -\frac{8 e \nu E_J \delta^{\scriptscriptstyle \Phi}}{3\hbar}
         \left(
         \begin{array}{ccc}
         2 &  -1 &  -1\\
         -1 &  2  &  -1\\
         -1 &  -1  &  2
         \end{array}
         \right);
         \label{Im22_pm}
\end{eqnarray}
the current operator $\hat{I}^{\rm\scriptscriptstyle (m1-1)}$
vanishes. Transforming to the basis states $|a,s\rangle$ of
$\sigma_x$, this takes the form
\begin{eqnarray}
    \hat{I}^{\rm\scriptscriptstyle (m3-3)}=
    -\frac{8 e \nu E_J \delta^{\scriptscriptstyle \Phi}}{3\hbar}
         \left(
         \begin{array}{ccc}
         3 &  0 &  0\\
         0 &  1  & -\sqrt{2}\\
         0 &  -\sqrt{2}  &  2
         \end{array}
         \right);
         \label{Im22_as}
\end{eqnarray}
obviously, the anti-symmetric state $|a\rangle$ is already a good
eigenstate of the current operator, while the states $|s\rangle$
and $|0\rangle$ remain mixed. Diagonalizing, we find the
eigenvalues $-8 e \nu \delta^{\scriptscriptstyle \Phi} j /\hbar$
and eigenvectors $|j\rangle$,
\begin{eqnarray}
             \begin{array}{c c}
              j \qquad & |j\rangle\\ \noalign{\vskip 6 pt}
             1 &  |a\rangle, \\
             \noalign{\vskip 4 pt}
             1  & [|s\rangle - \sqrt{2}|0\rangle]/\sqrt{3}, \\
             \noalign{\vskip 4 pt}
             0 & [\sqrt{2}|s\rangle + |0\rangle]/{\sqrt{3}}.
             \end{array}
             \label{jj}
\end{eqnarray}
The diagonal structure of $\hat{I}^{\rm\scriptscriptstyle
(m3-3)}$ in the subspace spanned by $|a\rangle$ and
$|s\rangle$ implies a measurement of the spin-component
$\sigma_x$ in the original basis $|\pm\rangle$.
Unfortunately, we have not been able to identify a setup producing
a complete diagonalization of the current operator; still, the
above constellation allows for a well defined statistical
procedure to identify the qubit state: Let us assume that just
before the measurement the wave function of the qubit belongs to
the doublet subspace, $|\psi\rangle = \psi_a|a\rangle +
\psi_s|s\rangle$, with $|\psi_a|^2+|\psi_s|^2=1$. The measurement
of the current eigenvalue $j$ then provides us with a
probabilistic measure of $P_s=|\psi_s|^2$: while the result $j=0$
appears with a probability $(2/3)\,P_s$, the current value $j=1$
is realized with a probability $(1/3)\,P_s + P_a = 1-(2/3)\, P_s$,
from which the desired quantity $P_s$ is easily derived within a
multi-shot measurement scheme. In an experiment tracing the Rabi
oscillations between the states $|a\rangle$ and $|s\rangle$ the
specific time dependence $P_s (t) = \cos^2(\Omega_{\rm Rabi} t)$
transforms into the probability difference ${\cal P}_{j=1}(t)
-{\cal P}_{j=0}(t) = 1/3-(2/3)\,\cos(2\Omega_{\rm Rabi} t)$; the
reduction in the amplitude is a consequence of the admixture of
the third state $|0\rangle$.

The actual measurement of the current is carried out with the same
(symmetric) measurement current configuration as discussed in the
previous section. The only difference is that in the present
measurement the current eigenvalues are proportional to $j=1$ and
$j=0$ instead of $\pm 1$, i.e., the switching of (one of the
two) junctions `n2-m1', `n3-m2' (or, depending on the
sign of flux bias, `n2-m3', `n3-m1') is characteristic
for a $j=1$ eigenstate, whereas the absence of any switching
corresponds to the $j=0$ state.

\section{Conclusions}
The CH$_4$ molecule can be viewed as a molecular analogue to 
the tetrahedral superconducting structure with the same 
tetrahedral symmetry group. The non-Abelian character of this
symmetry group is responsible for the natural appearance of
degenerate states. However, contrary to the situation in atomic
and molecular physics, where such degenerate levels usually
correspond to excited states, the macroscopic device discussed
here can be tuned such that the non-Abelian character of
its symmetry group manifests itself in the appearance of a
degenerate ground state. In order to do so, we have to bias the
device, both electrically and magnetically, to the maximally
frustrated point with half-flux $\Phi_0/2$ threading each sub-loop
and half-Cooper-pair charge $e$ induced on each island. 
The ground state is a doublet equivalent to a spin-$1/2$ 
system in a vanishing magnetic field.

Given the above analogy to molecular physics, one may ask whether
a splitting of the doublet ground state due to a Jahn-Teller type
instability may appear in our system as well. In the superconducting
tetrahedron this would correspond to a paramagnetic instability
with a spontaneous breaking of time reversal symmetry: paramagnetic
currents would lower the system energy due to their interaction
with the self-generated magnetic field and the ground state doublet
would split --- the magnitude of this effect is determined by the
magnetic inductance of the device which we have set to zero in our
analysis above. However, in order to realize such an instability,
the energy gain should be linear in the spontaneously generated
flux; since our device exhibits only a quadratic dependence on
flux, cf.\ (\ref{fieldsdd}), this type of instability is absent.

On a technical level, the superconducting tetrahedral qubit comes
with a number of practically useful features: {\it i)} The weak
quadratic sensitivity to electric and magnetic noise implies long
decoherence times for this device, similar to the `quantronium'
discussed by Vion {\it et al.} \cite{vion_02}. At the same time,
the quadratic dependence on bias allows for a manipulation via
$ac$-fields rather than the usual $dc$-bias, hence the most
dangerous low-frequency part of the noise spectrum can be blocked
from the qubit via appropriate filters. {\it ii)} The degenerate
ground state allows to avoid the appearance of phonon radiation
during idle time \cite{phon_rad}: Assume a qubit with states
$|0\rangle$ and $|1\rangle$ residing at different energies $E_0
\neq E_1$. A superposition state $|\Psi\rangle(t) = [\exp(-iE_0
t/\hbar)|0\rangle + \beta \exp(-iE_1 t/\hbar)
|1\rangle]/\sqrt{1+|\beta|^2}$ will induce voltage oscillations $V
= \hbar\dot\varphi/2e \propto (E_1-E_0)/2e$ across the Josephson
junction which couple to the underlying lattice via the
piezo-electric effect. Hence, the junction acts as an antenna
emitting phonons which contributes to the energy relaxation rate
of the qubit. In our tetrahedral qubit, $E_1 =E_0$ and this loss
channel is avoided during idle time. {\it iii)} The tetrahedral
qubit can be fabricated with junctions of relatively large size.
This is the consequence of a weak $\propto \exp[-{\rm const.}
(E_J/E_C)^{1/4}]$ rather than the usual $\propto \exp[-{\rm
const.} (E_J/E_C)^{1/2}]$ suppression of the qubit's operational
energy scale and entails two important advantages: first, less
stringent requirements on the fabrication process and better
junction uniformity, and second, an improved robustness of the
qubit with respect to charge noise originating from fluctuating
stray charges. The physical origin of this benevolent behavior is
found in the huge classical ground state degeneracy originating
from junctions with a simple $\propto \sin\varphi$ current-phase
relation combined with a maximal magnetic frustration; this
degeneracy then is lifted only due to quantum fluctuations
\cite{casimir_48, villain_80}. {\it iv)} The tetrahedral qubit
admits a large variety of manipulation schemes
--- arbitrary manipulations of the effective `spin $1/2$' ground
state can be implemented through either magnetic or electric bias
fields. {\it v)} The quantum measurement can be performed 
with respect to different basis states and using either
charge or flux bias. We have described schemes operating in the
qubit eigenbasis and projective measurements onto the
current basis. In the former setup, we have described detailed
procedures for the detection of both `spin' projections
$\sigma_z$ and $\sigma_x$. In the latter scheme, we have
identified a high-quality measurement scheme for the operators
$\sigma_z$ and $\sigma_y$ through appropriate charge bias; the
flux bias scheme produces a non-ideal but acceptable setup for the
measurement of $\sigma_x$. Proper rotation of the bias scheme by
$\pm 2\pi/3$ provides a measurement of the corresponding rotated
spin-components.

The above advantages seem worth the additional complexity of the
device. Still, one may pose the question whether the same benefits
can be implemented with a simpler device. E.g., the $C_{3v}$
symmetry group of the symmetric three-junction loop
\cite{ivanov_02} also contains a two-dimensional representation
and appropriate charging with $q_i=1/3$-charge per island produces
a doublet ground state suitable for quantum computation. However,
this simpler design does not exhibit the quadratic stability under
charge noise --- charge biasing one island reduces the symmetry to
$Z_2$ with only one-dimensional representations and the doublet
splits in linear order in $\delta^{\scriptscriptstyle Q}_i$. On
the contrary, charge biasing the tetrahedron reduces the symmetry
from $T_d$ to $C_{3v}$ which still contains a two-dimensional
representation. Indeed, the two complex conjugated states
$|\pm\rangle$ react the same way to a charge bias
$\delta^{\scriptscriptstyle Q}_i$ and hence the doublet splits
only in quadratic order. This behavior, the indifferent response
of the two ground state wave functions to a local perturbation is
strongly reminiscent of the idea of topological protection, where
the fault tolerance of the device is implemented on the hardware-
\cite{kitaev_97} rather than the software level
\cite{preskill_98}. The above symmetry arguments then show, that
in order to benefit from a protected degenerate ground state
doublet, the qubit design requires a certain minimal complexity;
it seems to us that the tetrahedron exhibits the minimal symmetry
requirements necessary for this type of protection and thus the
minimal complexity necessary for its implementation.

We acknowledge discussions with Ch.\ Helm, A.\ Ioselevich, P.\
Ostrovsky, and M.\ Troyer, and financial support from the Swiss
National Foundation (SCOPES, CTS-ETHZ), the Russian Ministry of
Science, the program `Quantum Macrophysics' (RAS), the RFBR grant
01-02-17759, and the NSF grant DMR-0210575. Numerical calculations
have been carried out on the Beowulf cluster ASGARD at ETHZ.


\end{document}